\documentclass[aps,a4paper,superscriptaddress,twocolumn]{revtex4-2}
\usepackage{tensor}
\usepackage{graphicx}
\graphicspath{ {images/} }
\usepackage[utf8]{inputenc}
\usepackage{amsmath}
\usepackage{amssymb}
\usepackage{enumerate}
\usepackage{subfigure}
\usepackage{tabularx}
\usepackage{lipsum} 
\usepackage[mathscr]{euscript}
\usepackage{float} 
\usepackage[colorlinks=true, pdfstartview=FitV, linkcolor=blue, citecolor=red, urlcolor=black, breaklinks=true]{hyperref}
\newcommand{\cmmnt}[1]{}
\newcommand{\be}{\begin{equation}}
	\newcommand{\ee}{\end{equation}}
\newcommand{\ben}{\begin{eqnarray}}
	\newcommand{\een}{\end{eqnarray}}
\newcommand{\bes}{\begin{subequations}}
	\newcommand{\ees}{\end{subequations}}
\def\bal#1\eal{\begin{align}#1\end{align}}
\newcommand{\etasc}{\scalebox{1.15}[1.15]{$\mathscr{\eta}$}} 

\newcommand{\sech}{{\rm sech}}
\newcommand{\LL}{{\mathcal L}}

\newcommand{\pu}{\mathrm{\partial_{\mu}}}

\newcommand{\Pu}{\mathrm{\partial^{\mu}}}

\newcommand{\qt}[1]{``#1''}
\newcommand{\pd}[2]{\ensuremath{\frac{\partial#1}{\partial#2}}}
\newcommand{\deriv}[2]{\ensuremath{\frac{d#1}{d#2}}}
\newcommand{\pb}[1]{\ensuremath{\partial_{#1}}}

\newcommand{\x}{\mathbf{x}}

\newcommand{\0}{\ensuremath{{\scalebox{0.6}{0}}}}
\begin{document}
	\title{Two-field models in the presence of impurities}
	
	\author{D. Bazeia}\affiliation{Departamento de F\'\i sica, Universidade Federal da Para\'\i ba, 58051-970 Jo\~ao Pessoa, PB, Brazil}
	\author{M. A. Liao}\affiliation{Departamento de F\'\i sica, Universidade Federal da Para\'\i ba, 58051-970 Jo\~ao Pessoa, PB, Brazil}
	\author{M. A. Marques}\affiliation{Departamento de Biotecnologia, Universidade Federal da Para\'iba, 58051-900 Jo\~ao Pessoa, PB, Brazil}
	
	\begin{abstract}
		
		This work deals with systems of two real scalar fields coupled to impurity functions, meant to model inhomogeneities often encountered in real physical applications. We investigate the theoretical properties of these systems and some of the consequences of impurity doping. We show that the theory may be modified in a way that preserves some BPS sectors, while also greatly impacting the behavior and internal structure of the solution, and exemplify those results with an investigation of a few interesting models in which impurities are coupled to a theory with a quartic potential. It is shown that, in the presence of impurities, the asymptotic behavior of field configurations may be changed, leading to solutions with different long-range properties, which are relevant to several physical applications. Our examples also highlight other important consequences that may follow from the addition of impurities, such as  the presence of zero-modes that can significantly change the internal structure of a given solution without altering its energy, the creation of new topological sectors that did not exist in the impurity-free theory, and the possibility of stable, nontrivial configurations generated by topologically trivial boundary conditions. We have also shown that it is sometimes possible to find energy minimizers in BPS sectors which were unpopulated in the canonical theory. These features show that impurities allow for significant flexibility in both the form of energy minimizers and the boundary conditions used to generate them, which may potentially broaden the range of applicability of the theory.
	\end{abstract}
	
	\maketitle
	
	\section{Introduction}\label{Intro}
	
	In field-theoretical investigations, one often deals with actions that are invariant under translation operations. This can be achieved through the use of an associated Lagrangian density lacking any explicit dependence on the spatial coordinates. This symmetry is related, through the celebrated Noether's theorem~\cite{Noether}, to the conservation of linear momentum, and amounts to the physical notion that all points of space should be \emph{a priori} equivalent in a fundamental theory. This is the setting in which many of the seminal works in topological field theory, such as~\cite{bogo, ps, Abrikosov, Nielsen, Hooft, polyakov, Rubinstein,SGIS, Skyrmion, skyrme}, have been conceived. 
	
	Despite those considerations, there are many  physical problems in which the hypothesis of translational invariance must be relaxed. One reason for this lies in the fact that the usual conservation theorems are valid for closed systems~\cite{goldstein}, which in real-life conditions amount to an idealization that can hold only approximately, to a degree that depends on the problem at hand. Translational symmetry may be spontaneously broken in the thermodynamic limit of condensed-matter theories~\cite{Altland}, and it may in fact not hold at all for effective theories, where inhomogeneities may be used to model imperfections in materials, external fields not accounted for in the Lagrangian, or the result of interactions with fixed background particles whose dynamics need not be included in a given theory. 
	
	The above considerations lead to the introduction of the so called impurities in field-theory, which effectively model the effects of inhomogeneity through the coupling with coordinate-dependent functions. Impurity theories may often be defined relatively to the homogeneous case. This can be achieved with use of an additive term that vanishes at regions in which homogeneity holds for a given material or background, and increases in regions where the interactions modeled  by these impurities are stronger. Physical consequences of the addition of impurities include the Anderson localization~\cite{Anderson, Andersonloc}, which happens when wave function scattering caused by impurities in a solid becomes strong enough to suppress wave propagation.  When impurities have magnetic moment, electron-impurity scattering may lead to the Kondo effect~\cite{Kondo, Sarachik, Matsuura, Tsay}, responsible for the non-monotonic relationship between electrical resistivity and temperature,  an observable effect even at low impurity densities. Impurities are also closely associated to engendering of the Quantum Hall effect and to many properties of the materials that display this effect~\cite{QHallI,QHallII,QHallIII,QHallIV}, as well as the associated quantum hall transitions~\cite{QHtransitions, QHtransitionsII}. Impurities can also be studied with the celebrated holographic correspondence~\cite{Maldacena}, and many investigations have been conducted in this manner~\cite{HolographyI, HolographyII, HolographyIII, HolographyIV, HolographyV, Kachru}.
	
	In recent years, several works have explored the effect of impurities on topologically nontrivial states, such as the topological vortices that are found in superconductors~\cite{Hook, Tong, PLB22, Ashcroft}, Bose-Einstein condensates~\cite{BEI, BEII, BEIII} and topological superfluids~\cite{SCI, SCII}. Other field-theoretical configurations with topological character that have been investigated in connection with impurities are the Skyrmions~\cite{Skyrmionimp, SkyrmionimpII}, nonlinear sigma solutions~\cite{O3}, emergent monopoles~\cite{Monopole, MonopoleII}, instantons~\cite{InstantonI, InstantonII} and dark solitons~\cite{DSI, DSII, DSIII}.
	
	Interactions between impurities and the kink solutions from topological one-field scalar theories have also been investigated for some time~\cite{Kink0, kinkI,KinkII, KinkIII, KinkIV, KinkV, Oscillons, AdamI, AdamII, kinks}, with many important discoveries such as different scattering properties, resonant states, modified long-range behavior and emergence of oscillon states. In ref.~\cite{AdamI}, the authors have verified that impurity coupling may be taken in a way that preserves half of the BPS sectors of the theory, generating a fundamental asymmetry between the kinks and antikinks of the model, and ensuring energy minimizers in the sectors in which the Bogomol'nyi bound can be saturated. The resulting model has been used to investigate some interesting phenomena that appear in this scenario, such as spectral walls~\cite{SpectralWallsI, SpectralWallsII}. In Ref.~\cite{BLM24}, this BPS description was extended to theories defined in generic spherically symmetric backgrounds, while in Ref.~\cite{BLMPLB} it was shown that stable nontrivial solutions can in fact be found even in higher dimensional flat spacetimes, as the impurities can be used to overcome the constraints imposed by the usual scaling theorems~\cite{Derrick, Hobart}.
	
	The above results for real scalar field theories were developed in the context of one-field models. This paper aims to extend these investigations to Lagrangian densities depending on two real scalar fields. In particular, we shall examine the possibility of BPS-preserving impurity doping in potentials that possess Bogomol'nyi bounds in the homogeneous scenario. The second and first-order equations will be derived in Section~\ref{gensec}, as well as the Bogomol'nyi bound for the pertinent topological sectors. Section~\ref{stability} is devoted to the derivation of the linear stability equations of the theory, formulated as an eigenvalue problem for the appropriate Hamiltonian. We confirm that solutions of the Bogomol'nyi equations are stable, and derive the zero-mode equations obtained from their linearization. We then proceed, in Section~\ref{examples}, to consider concrete models and calculate some of their solutions. To this end, we make use of the model introduced in~\cite{BNRT}, which has among its interesting features the BPS property, renormalizability and its rich vacuum manifold, which gives rise to multiple topologically nontrivial sectors in the homogeneous theory, with orbits of different form~\cite{BNRT5}. By coupling this model to various pairs of impurity functions, we find interesting properties and possibilities that did not exist in the canonical model. Finally, we summarize our main results in Section~\ref{Conclusions}, where perspectives of possible  applications and generalizations of our results are briefly discussed. 
	
	\section{General Theory and BPS equations}\label{gensec}
	
	In this work, we deal with two-field systems in two-dimensional Minkowski spacetime with metric tensor $\eta_{\mu\nu}=\rm{diag}(1,-1)$. We shall consider $\hbar=c=1$ and rescale the space and time coordinates, fields, impurities and parameters in order to use dimensionless quantities throughout the work. Methodologically, we consider a Lagrangian density of the general form
	\begin{equation}\label{Lag1}
		\LL=\LL_{\0} + f(\phi, \chi, \pb{x}\phi, \pb{x}\chi,x),
	\end{equation}
	where $\phi$ and $\chi$ are real scalar fields and 
	\begin{equation}
		\LL_{\0}=\frac{1}{2}\pu\phi\Pu\phi + \frac12 \pu\chi\Pu\chi - V(\phi,\chi),
	\end{equation}
	in which $V(\phi,\chi)$ is taken as a nonnegative function of the fields whose minima span a topologically nontrivial manifold, as usually required for the emergence of topological defects. We shall refer to $\LL_{\0}$ as the \qt{standard}, or homogeneous Lagrangian, because it generates the canonical equations of motion of a theory with two scalar fields.  It is readily seen that models governed by~\eqref{Lag1} are reduced to the standard two-field Lagrangian density $\LL_{\0}$ in the limit $f\to 0$. Thus, this function provides a measure of the extent by which a given system is deformed by the addition of impurities. This form is also convenient in that it allows for a direct comparison with the impurity-free Lagrangian $\LL_{\0}$, which exists as a limit in our theory.
	
	Although there is significant freedom in the choice of $f$, we restrict our attention to cases that allow for a BPS bound~\cite{bogo, ps}. We shall follow the procedure first carried out by Bogomol'nyi~\cite{bogo} to produce a topological lower bound for the energy in each sector. The ensuing bound is saturated by solutions of first-order equations  possessing the smallest energy compatible with given boundary conditions, thus providing a simple method through which energy minimizers may be found. In the standard theory, a BPS bound can be found provided that there exists an auxiliary function $W(\phi,\chi)$, sometimes called a superpotential in analogy with supersymmetry, such that
	\begin{equation}
		V(\phi,\chi)=\frac{1}{2}W_{\phi}^2+\frac12 W_{\chi}^2,
	\end{equation}
	where $W_{\phi}\equiv \partial W/\partial\phi$ and $W_{\chi}\equiv \partial W/\partial\chi$. In the presence of impurities, an additional restriction must be made to ensure the existence of a BPS bound. As we shall shortly verify, it suffices for this purpose to take $f(\phi, \chi, \pu\phi, \pu\chi,x)$ in the following manner:
	\begin{equation}\label{f}
		\begin{split}
			f&=\pb{x}\phi\sigma_1(x)+\pb{x}\chi\sigma_2(x)-W_{\phi}\sigma_1(x)\\
			&-W_{\chi}\sigma_2(x) -\frac12{\sigma_1^2(x)}-\frac12{\sigma_2^2(x)},
		\end{split}
	\end{equation}
	where the two last terms have been added for convenience to change the zero of the energy, as this sum has no effect in the variation of the action. Here, we work with two impurity functions $\sigma_1(x)$ and $\sigma_2(x)$, each coupled directly to one of the scalar fields and its derivatives. This generalizes the function $\sigma(x)$ introduced in~\cite{AdamI} for a theory with one real scalar field. One sees that $\LL\to \LL_{\0}$ in the limit $\sigma_1, \sigma_2  \to 0$. The introduction of these two functions takes into account the fact that $\phi$ and $\chi$ need not be equally affected by the source of inhomogeneity. It is possible that one of these functions is zero, as would be the case, for example, if $\LL$ is seen as an effective Lagrangian density derived from a theory in which some background field is coupled only to $\phi$. Subjecting the Lagrangian to first-order variations $\delta\phi$, $\delta\chi$ and imposing the condition $\delta S=0$, we are led to the Euler-Lagrange equations
	\begin{subequations}\label{EL}
		\begin{align}
			\begin{split}
				&\ddot{\phi}-\frac{\partial}{\partial x}\left(\pd{\phi}{x}-\sigma_1\right) + W_{\phi\phi}\left(W_{\phi} + \sigma_1\right)\\
				&+ W_{\chi\phi}\left(W_{\chi} + \sigma_2\right) =0,
			\end{split}\\
			\begin{split}
				&\ddot{\chi}-\frac{\partial}{\partial x}\left(\pd{\chi}{x}-\sigma_2\right) + W_{\chi\chi}\left(W_{\chi} + \sigma_2\right)\\
				&+ W_{\phi\chi}\left(W_{\phi} + \sigma_1\right) =0,
			\end{split}
		\end{align}
	\end{subequations}
	\noindent where overhead dots denote partial time derivatives. In order to allow for finite energy solutions, we subject Eqs.~\eqref{EL} to the boundary conditions
	\begin{align}\label{boundary}
		\lim_{x\to_{\pm\infty}}\left(W_{\phi} + \sigma_1\right)\to 0, && \lim_{x\to_{\pm\infty}}\left(W_{\chi} + \sigma_2\right)\to 0.
	\end{align}
	
	If these conditions are satisfied and the spatial derivatives of $\phi$ and $\chi$ fall to zero sufficiently fast, then the energy density of the system goes to zero asymptotically, which is a necessity for finite energy solutions.
	To completely specify the boundary conditions, the limiting values of the impurity functions need to be taken into account. Let
	\begin{align}\label{sigmainfty}
		&\lim_{x\to\pm\infty} \sigma_1(x)=A_{\pm}, && \lim_{x\to\pm\infty} \sigma_2(x)=B_{\pm},
	\end{align}
	where $A_{\pm}$ and $B_{\pm}$ are real constants. If both of these limits are zero, the impurity functions are localized, meaning that they are appreciably different from zero only at a finite neighborhood. In that case, the boundary conditions~\eqref{boundary} are independent of the sigma functions, and any finite energy solution must tend to an element of the vacuum manifold of $\LL_{\0}$. Throughout this work, we shall adopt the convention that a topological sector is completely specified by the boundary conditions~\eqref{boundary}, in such a way that two finite energy solutions belong to the same sector if, and only if, they can be continuously deformed into each other. This convention is adopted by some authors (such as~\cite{rajaraman}), but contrasts with part of the literature, in which sectors related by exchange of the indices are identified, so that, for example, a kink and antikink may be said to belong to the same sector. Neither convention is intrinsically better, but the one we adopted will prove adequate for our purposes, since impurities create a fundamental asymmetry between sectors with interchanged topological indices.
	\cmmnt{
		\begin{align}\label{ELALT}
			\begin{split}
				&\ddot{\phi}-\frac{\partial}{\partial x}\left(\pd{\phi}{x}-\sigma_1\right) + W_{\phi\phi}\left(W_{\phi} + \sigma_1\right)\\
				&+ W_{\chi\phi}\left(W_{\chi} + \sigma_2\right) =0
			\end{split}\\
			\begin{split}
				\text{and} \\ 	
				&\pu\left(\Pu\chi+\tensor{\delta}{^1_{\mu}}\sigma_2 \right) + W_{\chi\chi}\left(W_{\chi} + \sigma_2\right)\\
				&+ W_{\phi\chi}\left(W_{\phi} + \sigma_1\right) =0,
			\end{split}
	\end{align}}
	
	In Ref.~\cite{Hook}, it was demonstrated that it is often possible to introduce impurity coupling in such a way as to enable the preservation of half the BPS sectors of the theory. For a model with one scalar field, this was achieved in Ref.~\cite{AdamI} with a coupling function similar to that of~\eqref{f}. In our case, when $f$ is taken in this form one may complete squares in the energy functional of the theory to show that it can be written in the form
	\cmmnt{\begin{equation}\label{E}
			\begin{split}
				&E=\frac{1}{2}\int_{-\infty}^\infty dx \left[\left(\frac{\partial\phi}{\partial x}\right)^2+\left(W_{\phi}+\sigma_1\right)^2 -2\frac{\partial\phi}{\partial x}\sigma_1\right]+\\
				&\frac{1}{2}\int_{-\infty}^\infty  dx \left[\left(\frac{\partial\chi}{\partial x}\right)^2+\left(W_{\chi}+\sigma_2\right)^2 -2\frac{\partial\chi}{\partial x}\sigma_2\right]+T,
			\end{split}
	\end{equation}}
	\begin{equation}
		\begin{split}
			E=T +&\frac{1}{2}\int_{-\infty}^\infty\left(\pd{\phi}{x}-W_{\phi}-\sigma_1\right)^2 dx  \\
			+&\frac{1}{2}\int_{-\infty}^\infty\left(\pd{\chi}{x}-W_{\chi}-\sigma_2\right)^2 dx \\
			+&\int_{-\infty}^\infty\left(\pd{\phi}{x}W_{\phi} + \pd{\chi}{x}W_{\chi}\right) dx,
		\end{split}
	\end{equation}
	where $T=\frac{1}{2}\int dx [\dot{\phi}^2+\dot{\chi}^2]$ is the kinetic energy.
	Being equivalent to a surface term, the last integral is completely fixed by the boundary conditions, whereas every other term in the above functional is nonnegative. We thus conclude that the energy is subject to the bound
	\begin{equation}\label{EBPS}
		E\geq \Delta W \equiv  W(\phi(\infty), \chi(\infty)) - W(\phi(-\infty), \chi(-\infty)),
	\end{equation}
	with equality attained by configurations that satisfy the Bogomol'nyi equations
	\begin{subequations}\label{FO}
		\begin{align}
			&\dot{\phi}=\dot{\chi}=0 \\
			&\phi'=W_{\phi} + \sigma_1 \label{FOb} \\
			&\chi'=W_{\chi} + \sigma_2, \label{FOc}
		\end{align}
	\end{subequations}
	
	\noindent where the prime denotes partial differentiation with respect to $x$. The above equations give static solutions of~\eqref{EL}, as may be directly verified by differentiation. Solutions of these first-order equations are typically called BPS, in reference to Bogomol'nyi \cite{bogo}, and Prasad and Sommerfield~\cite{ps}. They are of great importance in theoretical physics because the Bogomol'nyi bound just derived ensures they are global minima of the energy functional, and hence classically stable. The topic of linear stability will be properly discussed in the following section.
	
	\section{Linear Stability Analysis}\label{stability}
	
	In this section, we will derive the differential equations that govern the linear stability of solutions of~\eqref{EL}. The procedure used to obtain these equations, which amounts to a linearization of the field equations, is well-known  in field theory, and typically leads to a  Schrödinger-like equation relating the time evolution of perturbations to the eigenvalues of an appropriately defined stability Hamiltonian. See, for example,~\cite{JG, rajaraman, Jackiw,Bazeiasantos,BBrito,Flores} for details. We then use the result to prove that solutions of~\eqref{FO} are stable under linear perturbations, aside from the usual zero-mode transformations. Although these results are to be expected, it is important to nevertheless perform this calculations because $(i)$  they have not yet, to our knowledge, been investigated for impurity-doped models of the form considered in this paper and $(ii)$ because the spectrum of the stability Hamiltonian determines the ground and excited modes of the solution, and thus plays a key role in defect dynamics, specially in scattering and the forces between defects  ~\cite{SugiyamaI, JPA54, PRL127, Gani}.
	
	In the interest of conciseness, we shall in this section introduce the notation $\{(\phi,\chi)\}\equiv \{(\varphi_{1},\varphi_2)\}$, which we use to write the field equations~\eqref{EL} in the form
	\begin{equation}\label{fieldeq}
		\Box{\varphi_{a}} +\sigma_a' +\frac{\partial ^2W}{\partial\varphi_{a}\partial\varphi_{b}}\left(\pd{W}{\varphi_{b}} + \sigma_b\right)=0,
	\end{equation}
	where $\Box\equiv \pu\Pu$ is the usual d'Alambert operator and summation is implied over repeated latin indices $a,b$, which would take the values $1,2$ for a two-field theory. Incidentally, this notation ensures that the analysis conducted throughout this section is valid for any number of fields, since~\eqref{fieldeq} is also formally valid for higher values of $a,b$. Since there is no extra cost to it, we might as well let $a,b=1,...,N$ for some integer $N$ for the remainder of this section. That extension amounts to a system of equations derived from a very straightforward $N$-field generalization of $\LL_{\0}$. In this notation, the BPS equations become 
	\begin{equation}\label{FOgen}\pd{\varphi_a}{x}=\pd{W}{\varphi_a} + \sigma_a. \end{equation}
	Equations~\eqref{fieldeq} can be linearized with the introduction of a first-order perturbation $\etasc_a$ that defines the transformation $\phi_a\to \phi_a+\etasc_a$. Introducing this perturbed form into~\eqref{fieldeq} and neglecting terms of order $(\eta ^2)$ and higher, we find, after grouping together equal powers of the perturbation and using~\eqref{fieldeq} to cancel-out the zeroth-order coefficients,
	
	\begin{equation}
		\Box\etasc_a + U_{ab}\etasc_b=0,
	\end{equation}
	where
	\begin{align}\label{Stabilitypot}
		U_{ab}=\pd{^2W}{\varphi_a\partial\varphi_c}\pd{^2W}{\varphi_c\varphi_b} + \pd{^3W}{\varphi_a\partial\varphi_b\partial\varphi_c}\left(\pd{W}{\varphi_{c}}+\sigma_{c}\right)
	\end{align}
	\noindent is the stability potential, and the derivatives of the superpotential are evaluated at the unperturbed solutions. Note that it differs from the analogous  potential of the homogeneous theory by an additive term consisting of a  linear combination of the impurity functions.
	
	This equation can be solved via separation of variables, which results in the Fourier decomposition
	\begin{equation}
		\etasc_{a}(t,x) = \sum_{k} \etasc_{a k}(x) \cos(\omega_k t),
	\end{equation}
	where the frequencies $\omega_k$ are determined by the eigenvalue equation
	\begin{equation}
		H_{ab} \etasc_{ak} = \omega_k^2 \etasc_{bk},
	\end{equation}
	where
	\begin{equation}\label{Hamiltonian}
		H_{ab} = - \delta_{ab}\deriv{^2}{x^2} + U_{ab}({x})
	\end{equation}
	is the stability Hamiltonian. Using~\eqref{Stabilitypot} and comparing~\eqref{Hamiltonian} with the analogous operator from the impurity-free theory, we can write
	\begin{equation}\label{18}
		H_{ab} = H_{ab}^{(0)} + \pd{^3W}{\varphi_a\partial\varphi_b\partial\varphi_c}\sigma_c, 
	\end{equation}
	where \cmmnt{$ H_{ab}^{(0)}=- \delta_{ab}\deriv{^2}{x^2} + W_{ac}W_{cb} + W_{abc}W_{c}$} $ H_{ab}^{(0)}$, which  can be inferred by comparing~\eqref{Hamiltonian} to~\eqref{18}, is formally identical to the stability Hamiltonian of the standard theory. This equation provides an elegant way to visualize, in matrix notation, the effect of impurity functions in the stability Hamiltonian of a theory with an arbitrary number of field equations of the form~\eqref{fieldeq}. It may also be useful if impurities are weak enough to allow for a perturbative approach in which $H_{ab}^{(0)}$ may be approximated by the impurity-free Hamiltonian.
	
	When the Bogomol'nyi bound is saturated, the first-order equations~\eqref{FOgen} can be used to factor this Hamiltonian into a quadratic form written as a product of first-order operators. This approach stems from supersymmetric theories~\cite{Cooper}, which are known to be closely related to BPS solutions~\cite{Witten}, and has been successfully applied to the standard theory in the past~\cite{Bazeiasantos}. In the present case, it is straightforward to verify that the introduction of the partner operators
	\begin{align}
		S_{ab}&=-\delta_{ab}\deriv{}{x} + \frac{\partial ^2W}{\partial\varphi_{a}\partial\varphi_{b}},\\ S_{ab}^{\dagger}&=\delta_{ab}\deriv{}{x} + \frac{\partial ^2W}{\partial\varphi_{a}\partial\varphi_{b}}
	\end{align}
	makes it possible to  write the stability equation in the form
	\begin{equation}
		S_{ab}^{\dagger} S_{ab}\etasc_{ak}=\omega_{k}^2\etasc_{ak}.
	\end{equation}
	
	Since the Hamiltonian is therefore seen to be a quadratic form, its spectrum must be devoid of negative frequencies, thus ruling out exponentially growing instabilities.
	
	One important subset among the solutions of the stability equation is given by the zero modes, which correspond to zero frequency solutions. These are related to transformations which change the field configuration in some way, but do not change its energy, thus being neutral in the sense of stability. These are very important in a dynamical setting, specially at small velocities, where the geodesic approximation can be used~\cite{geodesic, geodesicII}. For BPS (and hence static) solutions, the zero modes, and particularly their existence, are also important, as they indicate nontrivial transformations relating inequivalent solutions of~\eqref{FO} within a given topological sector. We shall see that these transformations result in more drastic changes to the internal structure of the fields when compared to the standard case. This richer structure is owed to the fact that the impurity functions are unchanged by these transformations, while the fields are not. 
	
	For solutions of~\eqref{FO1}, the zero modes must solve the system $S_{ab}\etasc_{ak}=0$. Explicitly,
	\begin{equation}\label{Zeromodegen}
		\deriv{\etasc_{ak}}{x}- \frac{\partial ^2W}{\partial\varphi_{a}\partial\varphi_{b}}\etasc_{bk}=0.
	\end{equation}
	
	Since this is a system of homogeneous linear first-order ODEs with as many unknown functions as equations, the usual existence theorems ensure that solutions will always exist under mild conditions. To be acceptable as a zero mode, it is however required that the solution is quadratic integrable, which does not follow automatically from the existence theorems. Indeed, this becomes an even more complicated issue in the presence of impurities, where such an analysis is best conducted in a case-by-case basis. In Ref.~\cite{AdamII},  a  thorough investigation of this issue has been conducted in the one-field analogue of~\eqref{Lag1}. In the aforementioned reference, the authors prove the existence of a generalized translational symmetry in the first-order equation, the action of which transforms the BPS kink in a distinct, but energetically equivalent, solution of the same sector, much in the same way as the usual translational symmetry relates all the static solutions in a standard theory. This symmetry is shown, in the same reference, to act trivially in states lacking a topological charge, which can be stable in theories with impurities. We expect a similar situation to occur here, and we do in fact find more than one BPS solution in topologically nontrivial sectors in the calculations we have performed. The topologically trivial sectors also agree  with the one field case in that they do not appear to possess any zero modes, as discussed in subsection~\eqref{lump}. 
	
	The identification of any existing zero modes with a generalized translation symmetry is however a far more complicated task in this case, as even in the $\LL_{\0}$ theory there may exist zero modes that cannot be identified with rigid translations of the system, since the moduli space generated by BPS equations can be two dimensional. Hence we shall for the moment leave the very interesting, but highly nontrivial, matter of identification of the symmetries responsible for the zero modes of this theory. We do however remark that the zero-mode equations can indeed be solved in many cases, as does in fact occur for many of the models we have examined in our investigations, some of which will be discussed  in detail in the next section. We shall see that the impurity functions engender far more complicated zero mode transformations, which give rise to a rich variety of solutions, with highly different features among themselves. 
	
	\section{Quartic potential models with impurtities}\label{examples}
	
	To advance our investigations, let us now commit to a definitive choice of $\LL_{\0}$ in order to consider concrete examples. The results developed in the previous section are general, and the  key features illustrated by the following examples can, in principle, be found for other choices of $V(\phi,\chi)$. It will however prove worthwhile to focus our efforts in a single choice of potential with multiple topological sectors. This means that we shall consider a family of Lagrangians in which each model is completely specified by the choice of impurity functions, thus allowing our results to be directly contrasted with the original model and with each other. This choice is meant to provide a clearer understanding of the deformations caused by different pairs of impurities in the theory, separating their effects from those that may ensue from changes of the potential itself.
	
	With this in mind, we shall take $\LL_{\0}$ as the Lagrangian density introduced in~\cite{BNRT}. The theory is obtained, for each choice of the free parameter $r>0$, through recourse to the auxiliary function
	\begin{equation}\label{WBNRT}
		W(\phi,\chi)=\phi(1-r\chi^2) - \frac{\phi^3}{3},
	\end{equation}
	from which the potential 
	\begin{equation}\label{potbnrt}
		V(\phi,\chi)=\frac{1}{2}(1-\phi^2 - r\chi^2)^2 + 2(r\chi\phi)^2
	\end{equation}
	can be derived. This model has been investigated in several subsequent works, see for example~\cite{BNRT2, BNRT3, BNRT4,BNRT5}, and references therein. Important applications include its use in Bloch Branes~\cite{BlochBrane, thick, degenetebranes} and nested domain defects~\cite{nested}. It has also been used to investigate Lorentz and CPT violations~\cite{LorentzviolationI,LorentzviolationII}, and to model Bifurcation and pattern changing in domain wall networks~\cite{bifurcation}. 
	
	This model has some important features that make it a natural choice. First, and perhaps of greatest theoretical importance, stands the fact that this model possesses a quartic potential, known to be the most general choice consistent with renormalization~\cite{QFT}. Moreover, this potential possesses sixteen topological sectors, twelve of which (or six, if the topological sectors are defined according to the conventions adopted in~\cite{BNRT, BNRT2}) are nontrivial. In a theory with impurities, where even the topologically trivial sectors may allow interesting stable solutions, leaves up to ten nonequivalent sets of boundary conditions that can be explored for different pairs of impurities. In Ref.~\cite{nested}, a reasonably general six  parameter cubic superpotential, defined as a sum of monomial terms with degree no greater than three, has been considered. The authors of that reference demonstrated that~\eqref{WBNRT} generates one of only two nonequivalent coupled models with $\text{Z}_2\times\text{Z}_2$ symmetry, being the one with more topological sectors among the two.

	The vacuum manifold of the theory generated by $\LL_{\0}$ is specified by the minima of this potential, which make up the set 
	\begin{equation}
		\{(\phi_{\0},\chi_{\0})\}=\{(\pm 1, 0), (0,\pm 1/\sqrt{r})\}.
	\end{equation}
	As previously mentioned, this manifold leads to twelve topologically nontrivial sectors, half of which lose the BPS property when impurities are included, as can be seen from the lack of $\pm$ signs in eq~\eqref{FO}. This is a well known feature of impurity-doped systems~\cite{AdamI,AdamII, Hook}. In a supersymmetric interpretation of the BPS equations, this property is explained by the fact that only half of the supercharges are preserved in this scenario~\cite{Hook}.

	If all impurities are assumed to be localized,  energy minimizers are still required to tend asymptotically to elements of the above set, even though $(\phi_{\0},\chi_{\0})$ is not itself a vacuum solution in the inhomogeneous scenario.
	
	The Bogomol'nyi equations~\eqref{FO} of this model are
	\begin{subequations}\label{FO1}
		\begin{align}
			\phi'&=1-\phi^2 -r\chi^2 + \sigma_1,\\
			\chi'&=-2r\chi\phi + \sigma_2.
		\end{align}
	\end{subequations}
	
	The zero-mode equations for this can be obtained directly from the linearization of the above equations, which lead to
	\begin{subequations}\label{zeromodeBNRT}
		\begin{align}
			&\deriv{\etasc_{1}}{x}=-2\left(\phi\etasc_{1} + r\chi\etasc_{2}\right),\\
			&\deriv{\etasc_{2}}{x}=-2r\left(\chi\etasc_{1} + \phi\etasc_{2}\right).
		\end{align}
	\end{subequations}
	As expected, these equations have the same form as those of the standard case, although their solution is of course different given that the fields (and hence the $W$ derivatives evaluated at the solution) are changed by the impurities.  As we shall now investigate in some detail, the long range behavior of solutions is changed by the presence of impurities which, as seen by the form of equations~\eqref{zeromodeBNRT}, completely determines the asymptotic properties of the zero-modes.
	
	\subsection{Asymptotic analysis}
	
	Let us now investigate the large $x$ behavior of some configurations. We assume the solution approaches $(\phi,\chi)=(0,1/\sqrt{r})$ as $x$ tends to infinity. For definiteness, let us subject~\eqref{FO1} to the boundary conditions
	\begin{subequations}\label{bc1}
		\begin{align}
			&\lim_{x\to_{-\infty}}\phi=-1, && \lim_{x\to_{-\infty}}\chi=0,\\
			&\lim_{x\to_{\infty}}\phi=0, && \lim_{x\to_{\infty}}\chi=\frac{1}{\sqrt{r}}.
		\end{align}
	\end{subequations}
	These conditions imply that solutions which saturate the Bogomol'nyi bound of this topological sector correspond to configurations with energy $E=\Delta W =2/3$.

	Using~\eqref{bc1} and~\eqref{FO1} one can write, for large positive $x$,
	\begin{align*}
		\phi=\psi +\mathcal{O}(2), && \chi=\frac{1}{\sqrt{r}} +\zeta + \mathcal{O}(2),
	\end{align*}
	where $\psi$ and $\zeta$ represent first-order deviations from the asymptotic values of the fields and powers of order two and higher have been neglected. Substituting these expansions in~\eqref{FO1} and neglecting terms of second-order or higher, one finds
	\begin{subequations}\label{AS1}
		\begin{align}
			\psi'&=-2\sqrt{r}\zeta + \sigma_1^{\infty} \label{AS1a} \\
			\zeta'&=-2\sqrt{r}\psi + \sigma_2^{\infty} \label{AS1b},
		\end{align}
	\end{subequations}
	where $\sigma_j^{\infty}$ stands for the behavior of the impurity near a point at infinity. Differentiation of~\eqref{AS1b} followed by substitution of~\eqref{AS1a} leads to
	\begin{equation}
		\zeta'' - 4r\zeta +2\sqrt{r}\sigma_1^{\infty}-\deriv{\sigma_2^{\infty}}{x}=0,
	\end{equation}
	which, coupled to the condition $\zeta(x\to\infty)=0$, gives the solution
	\begin{equation}\label{zetaas1}
		\begin{split}
			\zeta&= C_1e^{-2\sqrt{r}x} +\frac{1}{4\sqrt{r}}\left[e^{2\sqrt{r}x}\int_{1}^{x}e^{-2\sqrt{r}\xi}g(\xi)d\xi\right. \\
			&\left. + \  e^{-2\sqrt{r}x}\int_{1}^{x}e^{2\sqrt{r}\xi}g(\xi)d\xi\right], 
		\end{split}
	\end{equation}
	where
	\begin{equation}\label{g(xi)}
		g(\xi)\equiv \deriv{\sigma_2^{\infty}(\xi)}{\xi}-2\sqrt{r}\sigma_1^{\infty}(\xi)
	\end{equation}
	and $C_1$ is a real constant which must be chosen to match the behavior at the other endpoint of the spatial domain. By inserting~\eqref{zetaas1} into~\eqref{AS1a} one may easily obtain $\psi$ which, as seen from the form of~\eqref{AS1}, gives a similar behavior, namely
	\begin{equation}\label{psias1}
		\begin{split}
			\psi&= C_2e^{-2\sqrt{r}x} +\frac{1}{4\sqrt{r}}e^{2\sqrt{r}x}\int_{1}^{x}e^{-2\sqrt{r}\xi}h(\xi)d\xi \\
			& \ +\frac{1}{4\sqrt{r}}e^{-2\sqrt{r}x}\int_{1}^{x}e^{2\sqrt{r}\xi}h(\xi)d\xi,
		\end{split}
	\end{equation}
	where
	\begin{equation}\label{h(xi)}
		h(\xi)\equiv \deriv{\sigma_1^{\infty}(\xi)}{\xi}-2\sqrt{r}\sigma_2^{\infty}(\xi),
	\end{equation}
	with constant $C_2$. This illustrates the fact that \emph{both} impurity functions have, in general, an important effect in the asymptotic behavior of each of the scalar fields, even though only one of these functions appears in each of the Bogomol'nyi equations. The relative importance of each of the additive terms in eqs.~\eqref{g(xi)} and~\eqref{h(xi)}, as well as the order of approximation up to which they may be disregarded, depends of course on the form of the impurity. 
	
	In many cases, it may be possible to write asymptotic expansions of the form
	\begin{equation}\label{seriesexpansion}
		\sigma_i(x\to \infty) \sim \sum_{k} \frac{A_k}{x^k},
	\end{equation}
	which can be obtained by making the substitution $u=1/x$ and Taylor expanding the result. In that case, only the lowest-order terms in $g(x)$ and $h(x)$ need to be retained in an asymptotic expansion, so that we may consider expressions of the form $g(x)\sim 1/x^n$ in our calculations. Thus, the contributions of any given-order to~\eqref{psias1} and~\eqref{zetaas1} are determined by integrals of the form $\int e^{\pm2\sqrt{r}\xi}\xi^{-n}$. These can be solved analytically in terms of the so-called incomplete gamma functions~\cite{Arfken}. We may use these results to find formal expressions for $\psi$ and $\zeta$ as long as a nontrivial expansion of the form~\eqref{seriesexpansion} exists. The asymptotic behavior of such BPS solutions is thus of the form
	
	\cmmnt{We may thus use the fact that 
		\begin{align*}
			&\int_{1}^{x}\frac{e^{2\sqrt{r}\xi}}{\xi^n}d\xi=\left.r^{\frac{n - 2}{2}} \cdot 2^{n - 1} \left(-1\right)^{n} \operatorname{\Gamma}\left(1 - n,-2 \sqrt{r} \, \xi\right)\right|^{x}_1\\
			&\int_{1}^{x}\frac{e^{-2\sqrt{r}\xi}}{\xi^n}d\x=\left.-r^{\frac{n - 2}{2}} \cdot 2^{n - 1} \operatorname{\Gamma}\left(1 - n,2 \sqrt{r} \, \xi\right)\right|^{x}_1
	\end{align*}}
	
	\begin{equation}\label{asymptoticgen}
		\begin{split}
			\delta_i=-&\beta \frac{(4r)^{\frac{n-2}{2}}}{2} \left[ (-1)^ne^{-2\sqrt{r} x} \Gamma(1-n, -2\sqrt{r} x)\right. \\ &\left. + \ e^{2\sqrt{r} x} \Gamma(1-n, 2\sqrt{r} x) \right] +C_ie^{-2\sqrt{r}x}, 
		\end{split}
	\end{equation}
	where $\delta_1$ stands for $\zeta$ and $\delta_2$ for $\psi$.

	Assuming $n>1$ we may write, for large $x$,
	\begin{equation*}\label{Gamma}
		e^{\pm 2\sqrt{rx}} \Gamma(1-n, \pm 2\sqrt{rx}) \approx (\pm 2\sqrt{r}x)^{\left(\frac{1\mp 1}{2}-n\right) } ,
	\end{equation*}
	\cmmnt{
		\begin{equation*}\label{Gamma}
			e^{2\sqrt{rx}} \Gamma(1-n, 2\sqrt{r}x) \approx (2\sqrt{r})^{-n} x^{-n}
		\end{equation*}
		and 
		\begin{equation*}\label{Gammaa}
			(-1)^n e^{-2\sqrt{rx}} \Gamma(1-n, -2\sqrt{rx}) \approx (-1)^n (-2\sqrt{r})^{1-n} x^{1-n}
	\end{equation*}}
	which gives simple expressions that describe the asymptotic behavior of the field configurations in terms of integer powers of $1/x$. The above results hold as long as a convergent power series expansion of the form~\eqref{seriesexpansion} exists for the impurity functions of this model. 
	
	It may sometimes be the case that impurities decay exponentially or faster, corresponding to a situation in which every coefficient in expansion~\eqref{seriesexpansion} vanishes. This possibility will usually lead to an asymptotic behavior that is dominated by an exponential term, so that the fields will display a long-range behavior of the form  $\delta_i\sim e^{(-\mu x)}$, which is qualitatively similar to that of the standard theory, although the coefficient $\mu$, which plays a role roughly analogous to that of a mass parameter in a Yukawa-like potential, may be changed by the impurity. It is possible that  either zeros or singularities may appear all the way into infinity, thus preventing an expansion of the form~\eqref{seriesexpansion}. The former possibility will be considered in subsection~\ref{SecondModel}, where we see that well-behaved solutions are still found, with an asymptotic behavior that can be described as a combination of negative powers of $x$ and oscillatory functions.  Finally, one should note the theoretical possibility that the impurity function is chosen in such a way as to make one of the integrals in ~\eqref{zetaasympt} or \eqref{psias1} give a leading contribution that exactly cancels out the exponential term in these expressions, thus giving rise to a decay faster than the typical exponential behavior, such as a super-exponential tail. It seems, however, likely that this possibility would require some fine-tuning, so we shall not consider this scenario here.
	
	\subsection{First Model}\label{Firstmodel}
	
\begin{figure}[h]
		\centering
		\includegraphics[width=8.8cm]{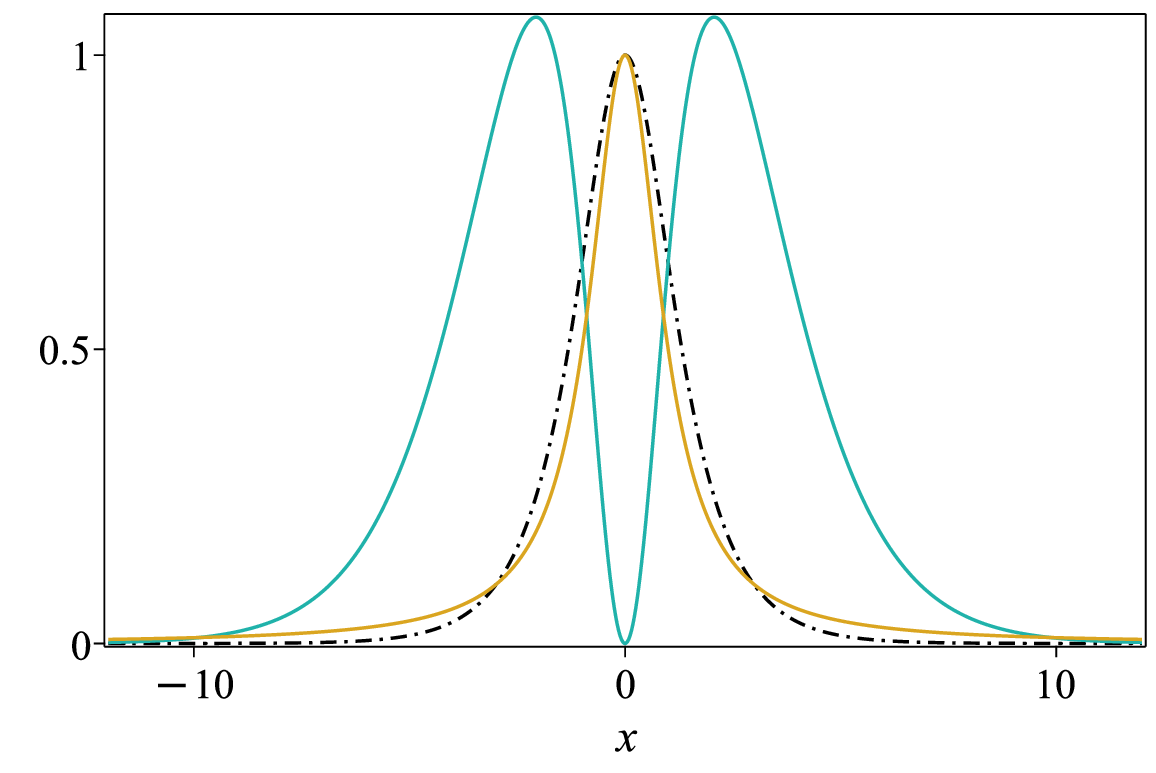} 
		\caption{Impurities $\sigma_1$ (black, dash-dotted line) and $\sigma_2$ (solid line, gold) from equation~\eqref{impex1} with $\alpha=\beta=1$ and $a=0$. The solid green line represents the impurity $\sigma_1=x^2\sech(x)$, corresponding to the case $\alpha=1$ and $a=2$.}
		\label{sigmaex1}
	\end{figure}

	Let us now explore a specific example of a solution satisfying the first-order equations with boundary conditions~\eqref{bc1}. To this end, consider the impurities 
	\begin{subequations}\label{impex1}
		\begin{align}
			&\sigma_1=\alpha x^a\sech(x),  \label{FeA} \\
			&\sigma_2=\frac{\beta}{1+x^2}, \label{FeB}
		\end{align}
	\end{subequations}
	where $a$ is a positive integer and $\alpha,\beta$ are real constants. The profile of these impurity functions can be seen in Fig.~\ref{sigmaex1}, for some specific values of these parameters.
	
	The asymptotic expansion of $\sigma_2(x)$ in powers of $1/x$ is given by
	\cmmnt{Given these choices, the BPS equations become
		\begin{subequations}\label{FOex1}
			\begin{align}
				&\phi'=\left(1-\phi^2-r\chi^2\right) + \alpha\sech(x)x^n\\
				&\chi'=2r\phi\chi + \frac{\beta}{1+x^2}.
			\end{align}
	\end{subequations} }
	\begin{equation}\label{expansion1}
		\frac{\beta}{1+x^2} = \frac{\beta}{x^2} - \frac{\beta}{x^4} + \mathcal{O}\left(\frac{1}{x^6}\right),
	\end{equation}
	whereas it is well known that the exponential decay of $\sech(x)$ is faster than any polynomial power, so that, regardless of the value of $a$, we may take $\sigma_1\sim 0$ asymptotically, at any finite order approximation.
	
	We may insert the leading order term of~\eqref{expansion1} into~\eqref{psias1} to find 
	\begin{equation}\label{zetaasympt}
		\psi(x)\approx C_2 e^{-2\sqrt{r}x} + \frac{\beta}{2\sqrt{r}x^2},
	\end{equation}
	which decays as an inverse square function at leading order, much slower than the exponential decay that one would have found in the absence of impurities. For the function $\zeta$, what is relevant is the derivative of $\sigma_2^{\infty}$, which satisfies
	\begin{equation}
		\deriv{\sigma_2^{\infty}}{x}\approx \beta\left(-\frac{2}{x^3} + \frac{4}{x^5}\right)
	\end{equation}
	at large $x$. When combined with~\eqref{zetaas1}, this result leads to an integral equation that we can solve to obtain
	\begin{equation}
		\zeta(x) \approx C_2 e^{-2\sqrt{r}x} + \frac{\beta}{2rx^3},
	\end{equation}
	which behaves as $1/x^3$ at large distances for any nonzero $\beta$. Thus, the same impurity has been responsible for the change in the tail of the scalar fields that make up this configuration. It produces a different effect on each field, so that, unlike what was seen in the absence of impurities, they fall to their asymptotic values at different rates. 
	
\begin{figure}[h]
		\includegraphics[width=8.5cm]{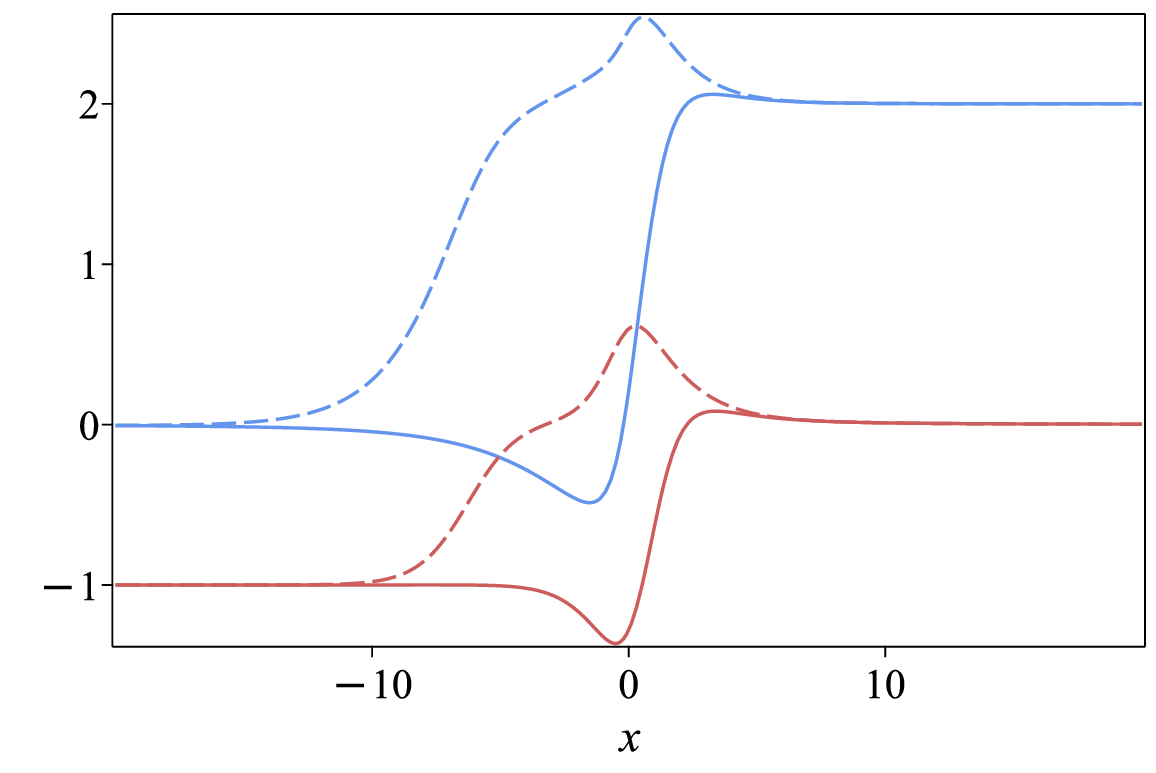}
		\caption{Solutions of~\eqref{FO1} with boundary conditions~\eqref{bc1} and impurities ~\eqref{impex1} with $\beta=1$, $r=\frac{1}{4}$, and $a=0$. Here, $\phi$ is represented as a red line and $\chi$ as a blue one. The solid and dashed lines represent two different solutions connected by zero modes.  For the solid line solution, we have, to three significant digits,  $(\phi(0),\chi(0))=(1.28,0.211)$, while $(\phi(0),\chi(0))=(0.597,2.46)$ with the same rounding.  }
		\label{ex1fields}
	\end{figure}

	The full equations~\eqref{FO1} cannot be solved in closed-form, but a numerical evaluation is possible, and has been conducted. The results are displayed in Fig.~\ref{ex1fields} for the choices $\alpha=\beta=1$ and $n=0$, in which two different BPS solutions satisfying~\eqref{FO1} are depicted. Both solutions possess the same limiting values and energy, and are thus related by zero-mode transformations, but we see that the differences between these solutions are far more significant than rigid or relative translations between the fields. Note that their internal structure is strikingly different. Indeed, both solutions are composed of functions possessing critical points near the impurity centers, but these are of opposing concavity, as the fields in the solid line solution possesses minima, while in each of the dashed lines a maximum is found. Away from the impurities, the fields converge to the same asymptotic behavior.
	
	Note that, in both of the solutions shown in Fig.~\ref{ex1fields}, $\phi$ and $\chi$ display the appropriate asymptotic behavior, but, unlike what happens in the analogous configuration of the impurity-free model, none of them is a monotonic function. This happens because, in the region where the $\sigma$ differ considerably from zero, the coupling with impurities may be stronger than the one induced by the potential, thus temporarily driving the fields away from the minima of $V(\phi, \chi)$.
	
\begin{figure}[h]
		\centering
		\includegraphics[width=8.8cm, trim=0.8cm 0 0 0]{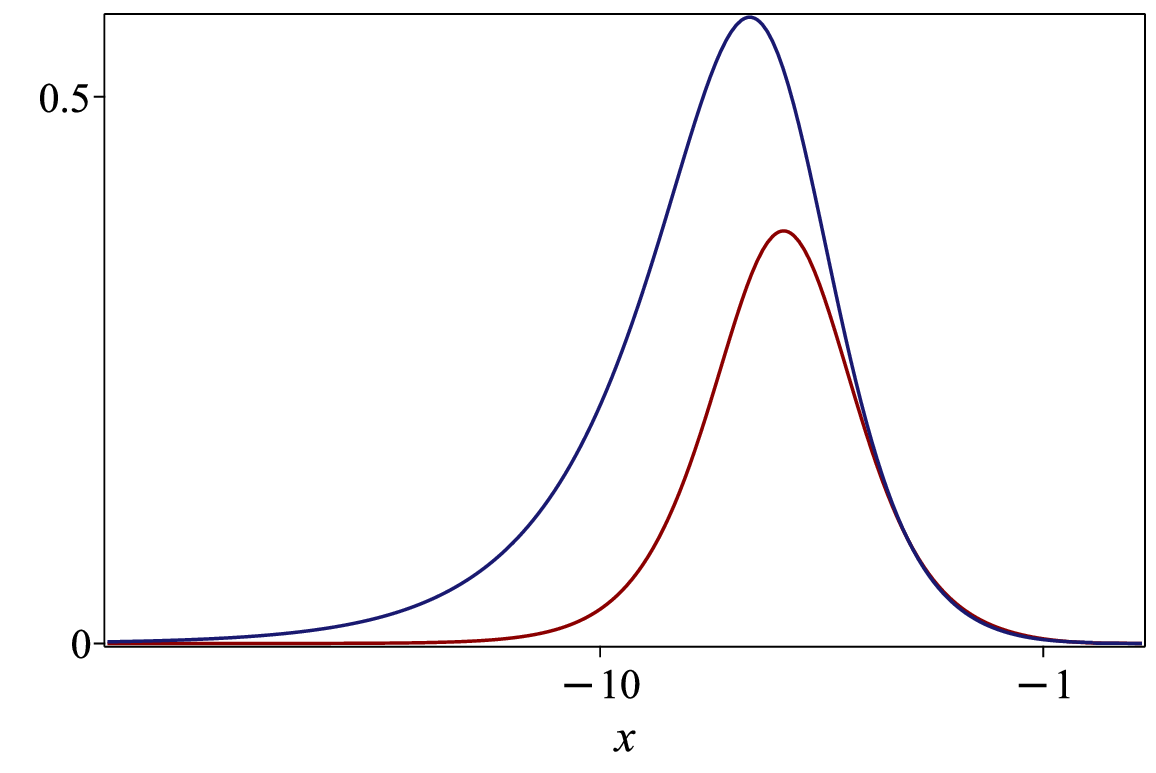}
		\caption{Zero-mode solving~\eqref{zeromodeBNRT} with $(\phi,\chi)$ obtained as a first-order perturbation from the dashed line solution depicted in Fig.~\ref{ex1fields}. Here, the dark red line represents the $\eta_1$ perturbation, while the dark blue one corresponds to $\eta_2$. }
		\label{zeromode} 
	\end{figure}

	Equations~\eqref{zeromodeBNRT} can in fact be solved for the system considered in this subsection to find normalizable zero-modes. One such solution, obtained from the perturbation of the dashed-line solution from Fig.~\ref{ex1fields}, is depicted in Fig.~\ref{zeromode}, where two asymmetrical lump profiles are seen. The fact that acceptable solutions of the zero-mode equations exist is consistent with the presence of more than one solution in this topological sector, as discussed above and shown in Fig.~\ref{ex1fields}. Zero modes can in fact be found for many of the examples we shall consider throughout this work, although we shall not pause to discuss these modes in all of the subsequent examples, instead choosing to focus on other aspects of the configurations, and including discussions of zero modes in some relevant cases. 
	
\begin{figure}[h]
		\centering
		\includegraphics[width=8.9cm]{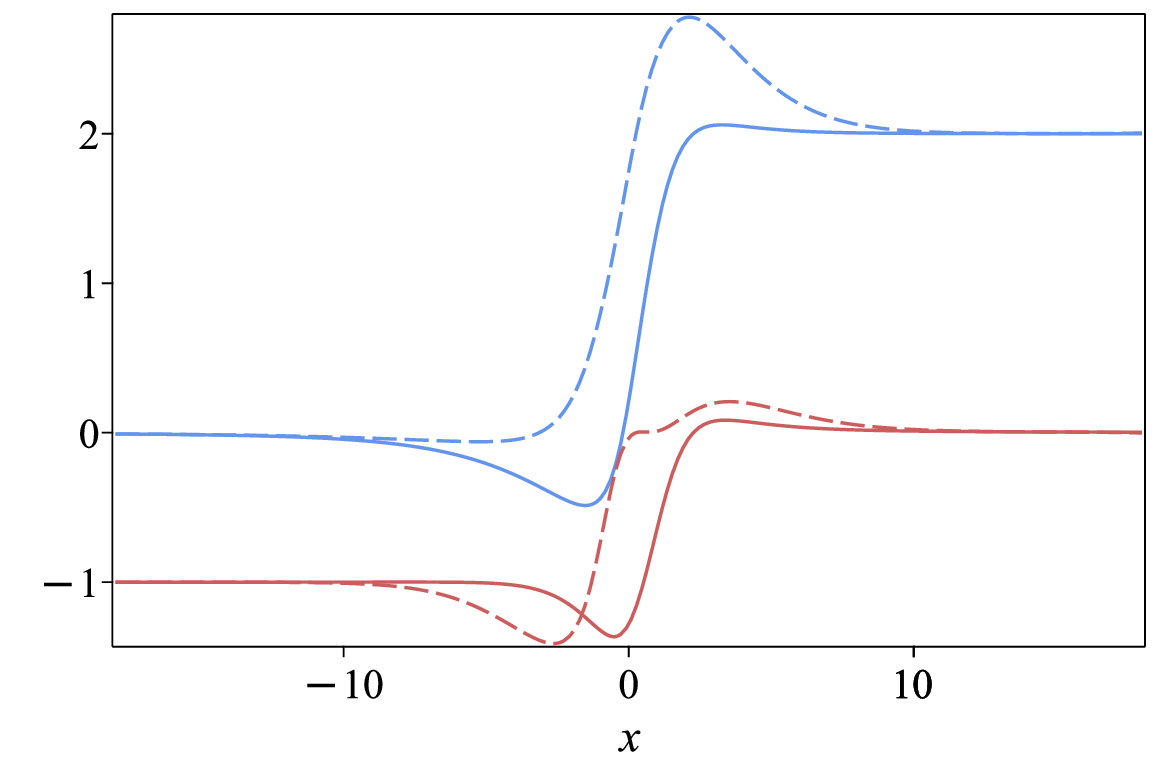}
		\caption{Solution of~\eqref{FO1} with boundary conditions~\eqref{bc1} and impurity functions ~\eqref{impex1} with $\beta=1$, $r=\frac{1}{4}$, and $a=2$. Here, $\phi$ is represented as a red line and $\chi$ as a blue one. For the solid line solution, we have $(\phi(0),\chi(0))=(-1.28,0.211)$, while $(\phi(0),\chi(0))=(-0.0356,1.75)$ to three significant digits.}
		\label{ex1.5}
	\end{figure}

	As previously stated, the value of $a$ in equation~\eqref{FeA} does not play a significant role in the asymptotic behavior of BPS configurations. However, this power of $x$ does play a significant role in a neighborhood of the origin, as it can in fact engender significant changes in the profile of the impurity and therefore in the fields themselves. In Fig.~\ref{sigmaex1}, the impurity~\eqref{FeA}, with $a=2$, is displayed as dashed line. We see that, unlike the previous example, this impurity now has two symmetrically placed maxima rather than one, and vanishes at the origin. Solutions obtained for the system of equations engendered by this choice of $a$ are displayed in Fig.~\ref{ex1.5}, where we again depict two solutions that solve these equations in the same topological sector, which, as in the previous example, display, in an interval containing the origin, significant structural differences in relation to each other.

	\subsection{Second model}\label{SecondModel}
	As a second example, let us investigate impurities with a more complicated asymptotic behavior. Let
	\begin{subequations}\label{Impex2}
		\begin{align}
			&\sigma_1=\alpha \frac{J_2(x)}{x^a}, \\
			&\sigma_2=\beta \frac{J_2(x)}{x^b},
		\end{align}
	\end{subequations}
	where $J_{2}$ denotes a Bessel function of the first kind and order 2, and $a,b$ are positive constants. To avoid singularities, these parameters must be constrained. Indeed, near the origin, Bessel functions whose order is not a negative integer are known to satisfy~\cite{NISTI}
	\begin{equation}\label{Bessel}
		J_{\nu}(x)=\left(\frac{x}{2}\right)^{\nu}\frac{1}{\Gamma(\nu+1)} + \mathcal{O}(x^2),
	\end{equation}
	so that the impurity functions of the form shown in in~\eqref{Impex2} behave as $\sigma_i\sim x^{2-n}$, where $n=a,b$. Hence, a singularity at the origin is avoided provided  $a,b\leq2$. The allowed integer values for these parameters are thus $a,b=0,1,2$, leading to nine such combinations. If an arbitrary Bessel function of an order other than two is used in the definition of~\eqref{Impex2}, then this inequality is generalized to $a,b\leq\nu$, which may be used as a tool for applications requiring a different asymptotic behavior of the fields. 
	
\begin{figure}[h]
		\centering
		\includegraphics[width=8.8cm, trim={1cm 0 0 0},clip]{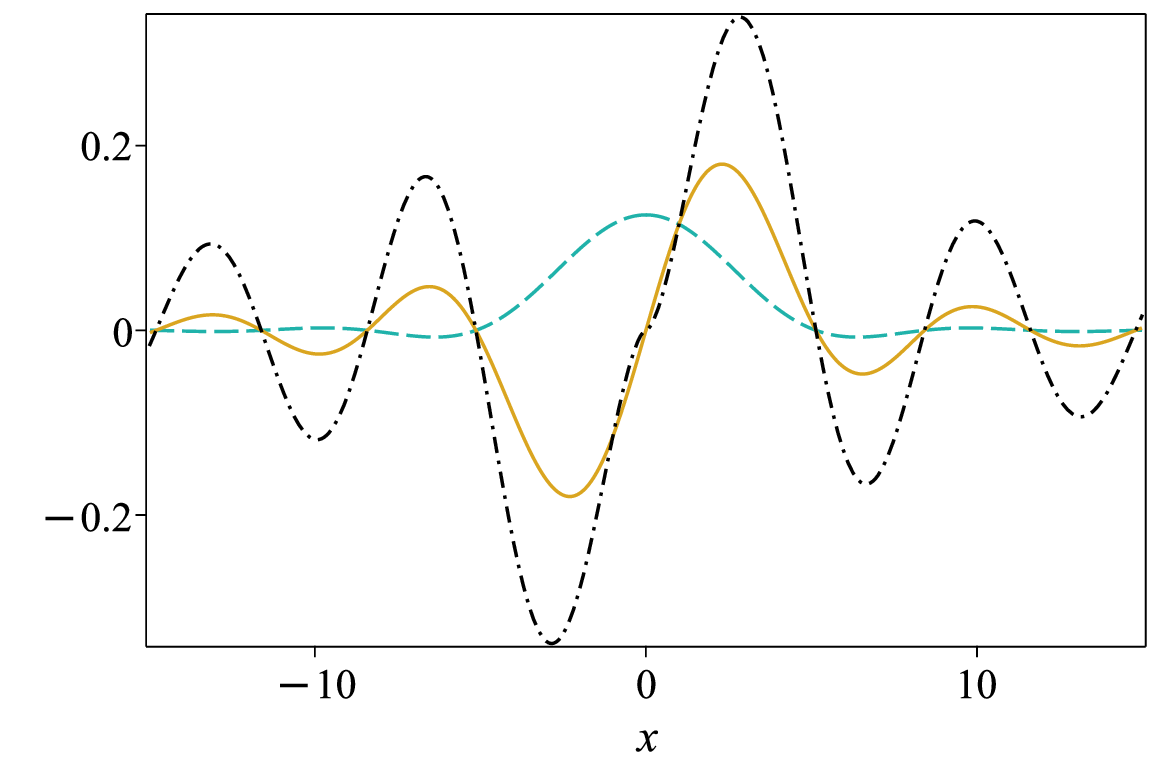} 
		\caption{Impurities $J_2(x)/x^{n}$ for $n=1$ (dashed green line), $n=2$ (solid gold line) and $n=1/3$ (black dash-dotted line). The case $n=1/3$ has been extended to allow for negative values of $x$, which was achieved by selecting the real root of $x^{n}$, rather than the principal root. For the other values of $n$ depicted, both roots coincide. }
		\label{sigmaex2}
	\end{figure}

	Note that it is not necessary to take $a$ and $b$ as integers in~\eqref{Impex2}. However, care must be taken to assure that the impurities remain real. If the principal root of the power function $x^{n}$, for fractional $n$, is used, then $\sigma_i$ must be of the form~\eqref{Impex2} only for $x>0$, with $\sigma(x<0)=0$. However, it is also possible in some cases (as is the case, for example, when $n$ is the reciprocal of an odd number) to find a real root for negative values of $x$. In such scenarios, we may extend the domain of $\sigma$ by selecting the real root of $x^{n}$. This is equivalent to writing $\sigma_i\propto \frac{J_2(x)}{(-|x|)^n}$ for $x<0$, and then defining the function piecewisely. This extension corresponds to the negative values of the dash-dotted line in Fig.~\ref{sigmaex2}. At positive $x$, the principal root itself is real, so these fractional powers can be seamlessly included in the asymptotic analysis we shall now perform.

	Due to the fact that Bessel functions still oscillate at arbitrarily large values of $x$, a true asymptotic expansion of the form~\eqref{seriesexpansion} is not possible for these impurities. However, a limiting form consistent with the decaying oscillation amplitude of these functions can be found. For the functions at hand we can write, at large $x$~\cite{NISTI},
	
	\begin{equation}\label{impex2}
		\frac{J_2(x)}{x^{\gamma}} \approx -\sqrt{\frac{2}{\pi}} \left( \frac{\sin\left(x + \frac{\pi}{4}\right)}{x^{\frac{1}{2}+\gamma}} + \frac{15 \cos\left(x+ \frac{\pi}{4}\right)}{8 x^{\frac{3}{2}+\gamma}} \right),
	\end{equation}
	where $\gamma$ is a real constant. Hence, the tail of each of the scalar fields of a configuration doped with impurities~\eqref{Impex2} is expected to oscillate around the asymptotic values with a decaying amplitude, with a rate of decay controlled by the parameters $a$ and $b$ of the model. Both terms in the above expansion need to be considered in the asymptotic analysis of the solutions, since the dominant power in that expansion may change when the zeroes of $J_{2}(x)$ are met.
	
\begin{figure}[h]
		\centering
		\includegraphics[scale=0.45, trim={1.2cm 0 0 0},clip]{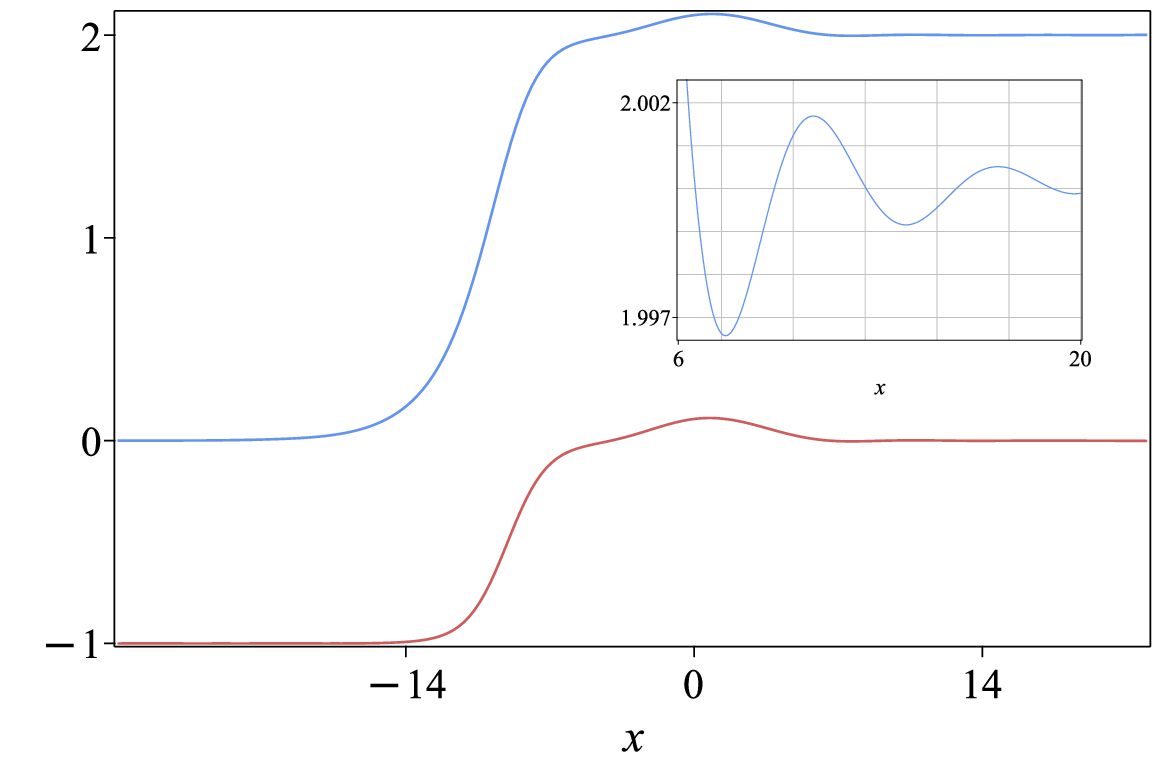}
		\caption{Solution of~\eqref{FO1} with boundary conditions~\eqref{bc1} and impurities~\eqref{Impex2}, with $\alpha=\beta=1$, $a=b=2$ and  $r=\frac{1}{4}$. Here, $\phi$ is represented as a red line and $\chi$ as a blue one. The subplot within this figure is a zoomed in graph where the asymptotic oscillations of $\chi(x)$ can be seen more clearly. Similar oscillations are found in $\phi(x)$.}
		\label{Fig2}
	\end{figure}

\begin{figure}[h]
		\centering
		\includegraphics[width=8.1cm,trim={1.2cm 0 0 0}, clip]{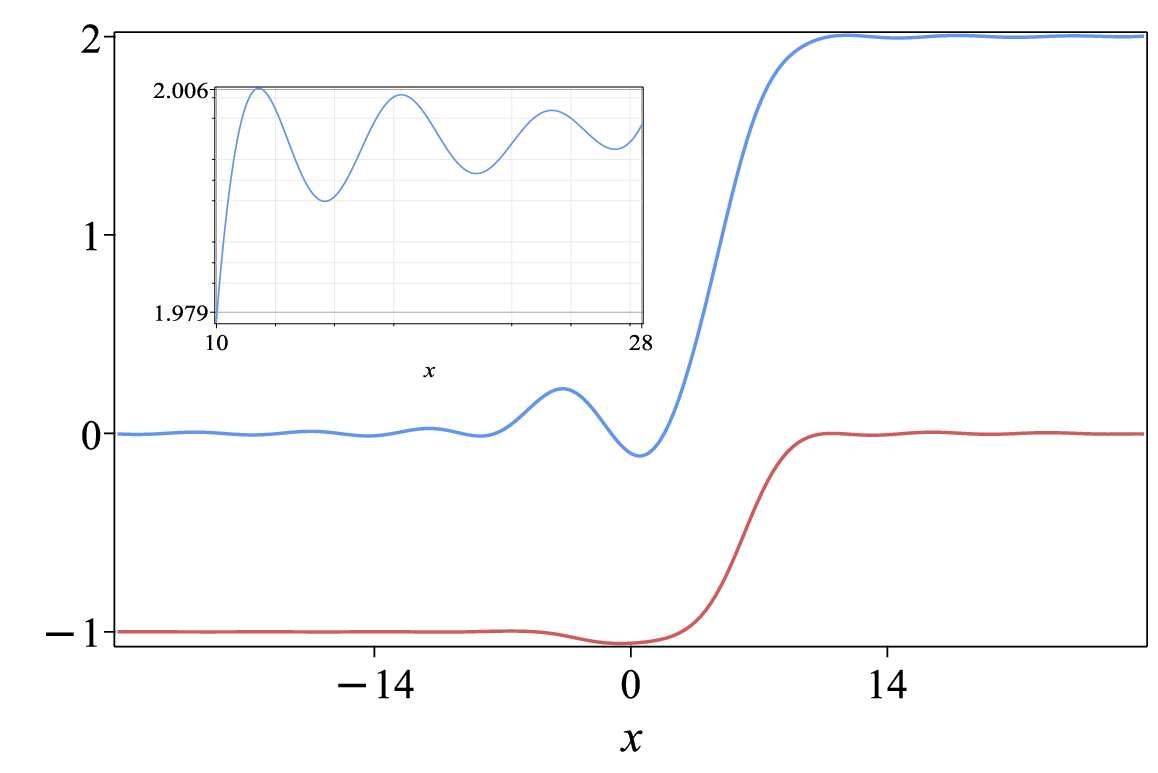}
		\caption{Solution $\phi$ (red), $\chi$ (blue) of~\eqref{FO1}  with impurities~\eqref{Impex2}, subject to boundary conditions~\eqref{bc1}. The parameters of the model are taken as $\alpha=\beta=1$, $a=2$, $b=1$ and  $r=\frac{1}{4}$. The subplot within this figure is a zoomed in graph where the asymptotic oscillations of $\chi(x)$ can be seen more clearly. Similar oscillations are found in $\phi(x)$.}
		\label{Fig3}
	\end{figure}

	As before, we may use equations~\eqref{zetaas1} and~\eqref{psias1} to write $\zeta$ and $\psi$ analytically in terms of incomplete Gamma Functions for any given choice of $a$ and $b$. In the large $x$ limit, we may use~\eqref{impex2} to show that the integral  expressions in these asymptotic expansions become linear combinations of contributions of the form
	\begin{equation}\label{Besselsineint}
		\begin{split}
			&e^{\mp2\sqrt{r}x}\int \frac{\sin\left(x + \frac{\pi}{4}\right) e^{\pm2\sqrt{r}x} \, }{x^n} \, dx \approx \left(\frac{i\pm 1}{\sqrt{2}}\right)\frac{e^{\mp ix}}{x^{n}} \Biggl\{\\ &\hspace{25pt}\frac{1}{2\left(2\sqrt{r}-i\right)} \left[ 1 \pm \frac{n}{2 \left(2\sqrt{r}-i\right) x} \right]\Biggr\} + \text{C.C}
		\end{split}
	\end{equation}
	and 
	\begin{equation}\label{Besselcosint}
		\begin{split}
			&e^{\mp2\sqrt{r}x}\int \frac{\cos\left(x + \frac{\pi}{4}\right) e^{\pm2\sqrt{r}x} \, }{x^n} \, dx \approx \left(\frac{i\pm 1}{\sqrt{2}}\right)\frac{e^{\pm ix}}{x^{n}} \Biggl\{\\ &\hspace{25pt}\frac{1}{2\left(2\sqrt{r}+i\right)} \left[ 1 \pm \frac{n}{2 \left(2\sqrt{r}+i\right) x} \right]\Biggr\} + \text{C.C},
		\end{split}
	\end{equation}
	
	\noindent where $n$ is a positive constant equal to $a,b,a-1$ or $b-1$ as derived from substitution of~\eqref{impex2} into~\eqref{g(xi)} and \eqref{h(xi)}, while $i$ is the imaginary unit  and C.C denotes the complex conjugate of the terms preceding it. Because these expressions are sums of pairwise conjugate functions, they are always real, and in fact amount to leading contributions that consist of products between (co)sine functions and inverse  powers of $x$. Since~\eqref{Besselsineint} and~\eqref{Besselcosint} are out of phase in relation to each other, their zeros cannot coincide, and thus one of these expressions always dominates the exponential terms in~\eqref{zetaas1} and~\eqref{psias1}. This leads to damped oscillations with a rapidly falling amplitude, corresponding to leading term of the form
	\begin{equation}
		\delta_i\sim x^{-n}\left( A\sin(x) + B\cos(x)\right),
	\end{equation}
	where $A$ and $B$ are constants depending on $r$ and $n$, with an error whose first correction depends on $\left(1/x^{n+1}\right)$.  This behavior is exemplified by the $\chi(x)$ field in the subplots drawn in Figs.~\ref{Fig2} and~\ref{Fig3}. Note that both of these subplots have been made at very significant distances of the origin, close to which both the defects and impurities are centered. And yet, even at these distances, the damped oscillations remain noticeable after two or three decimal digits, which are still measurable at reasonable energy scales. These amplitudes can be made greater if the $\alpha$ and $\beta$ parameters in~\eqref{Impex2} are increased.
	
	Let us now deal with full the Bogomol'nyi equations of this problem. Solutions have again been numerically found for the relevant boundary conditions, and the results can be seen in Fig.~\ref{Fig2} for the case $a=b=2$ and in Fig.~\ref{Fig3} for the case $a=2$, $b=1$. The damped oscillations predicted from our asymptotic analysis are shown in these Figures, with a faster amplitude decay of the $\chi$ field observed in the first example, where the oscillations have almost disappeared at the rightmost extreme of the plot. This results from the fact that $b=2$ in this example, while this parameter was set to unity in the second one. Since the two impurities used are of the same form, a similar graph for the $\phi$ field can be obtained if the relevant $y-$axis region is zoomed in.

	\cmmnt{
		\begin{subequations}\label{FOex2}
			\begin{align}
				&\deriv{\phi}{x}=\left(1-\phi^2-r\chi^2\right) + \alpha \frac{J2(x)}{x^a}\\
				&\deriv{\chi}{x}=2r\phi\chi + \beta \frac{J2(x)}{x^b}
			\end{align}
	\end{subequations}}

	\subsection{Third model}
	
	We may also investigate other topological sectors in the systems explored in the previous examples. As previously mentioned, the potential~\eqref{potbnrt} gives rise to six different topological sectors, three of which possess the BPS property. We may thus explore the behavior of solutions in other BPS sectors, which may be achieved by the consideration of different boundary conditions. One interesting possibility is 
	\begin{align}\label{bc2}
		&\lim_{x\to_{\pm\infty}}\phi=0, && \lim_{x\to_{\pm\infty}}\chi=\pm\frac{1}{\sqrt{r}}.
	\end{align}
	These boundary conditions also define a valid topological sector in the impurity-free theory, but this is a non BPS sector, in the sense that, although the second-order equations can indeed be solved to derive finite energy configurations in this sector, any such solution gives rise to a total energy strictly above the Bogomol'nyi level. This feature is a consequence of the fact that~\eqref{bc2} and~\eqref{EBPS} imply $E=\Delta W=0$ for this sector, a value only achievable by trivial solutions in the standard model. Such configurations do not, of course, satisfy the boundary conditions~\eqref{bc2}.  However, it is well known that, for impurity-doped systems, the energy density has contributions which need not be positive at all points~\cite{AdamI, Hook, Ashcroft}. This may be explained by the interpretation of~\eqref{Lag1} as an effective Lagrangian in which the impurity functions correspond to the interaction of $\phi,\chi$ with external fields whose potential energy is not taken into account by~\eqref{Lag1}. In any case, the possibility of negative energy density in some regions means that boundary conditions such as~\eqref{bc2} may give rise to nontrivial solutions which are stable despite the zero BPS bound.
	
	It is clear that the existence of the solutions discussed above is dependent on the choice of impurity functions, which must allow for a zero to emerge in the right-hand side of ~\eqref{FOb} and must be compatible with the kink-like behavior that $\chi$ must display under these boundary conditions. However, it is not too difficult to come up with functions with the appropriate behavior. In fact, we may even find closed-form solutions in some cases. One very simple choice that allows for configurations with the desired behavior is given by
	\begin{subequations}\label{impurityex3}
		\begin{align}
			\sigma_1(x)&=-\sech(x)\tanh(x),\\
			\sigma_2(x)&=\left(2+\sinh(x)\right)\sech^2(x).
		\end{align}
	\end{subequations}
	\begin{figure}[h]
		\centering
		\includegraphics[width=8.8cm]{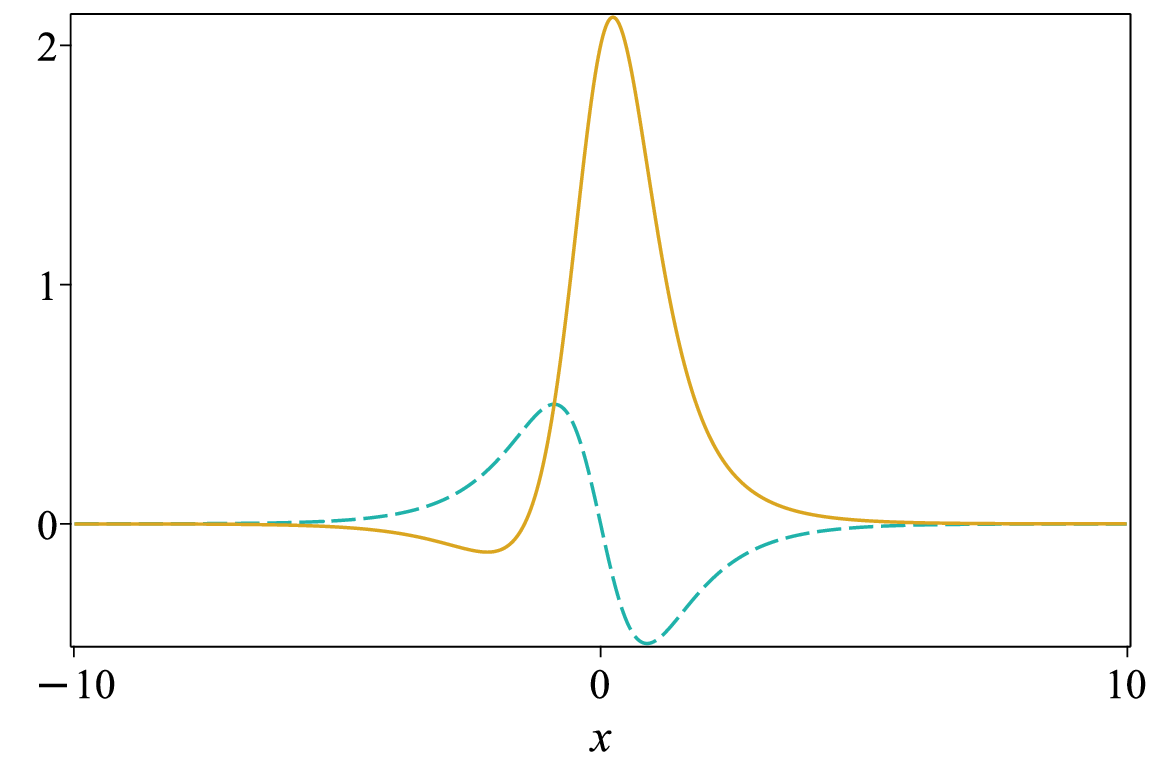}
		\caption{Impurities $\sigma_1$ (green, dashed) and $\sigma_2$ (yellow) from equation~\eqref{impurityex3}.}
		\label{sigmaex3}
	\end{figure}  
	The behavior of these impurity functions can be seen in Fig.~\ref{sigmaex3}. We note that $\sigma_1$ is an odd function which falls rapidly from its maximum in a neighborhood of $x=0$, crossing the $x$ axis at the origin, while $\sigma_2$ is strictly positive and has a peak in that same neighborhood. This is consistent with a lump-like $\phi$ field with a maximum at the origin. In this case, the height of this maximum stands at $\phi_{max}=\sqrt{1 -r\chi^2(0)}$ and ultimately depends on the value of $\chi$ at this point.  If, in particular, $\chi(0)=0$, then $\phi_{max}=1$.  If $r=1/4$. Namely, the first-order equations are solved by 
	\begin{subequations}\label{sol3x3}
		\begin{align}
			\phi(x) &=\text{sech}(x),\\
			\chi(x) &= 2\tanh(x).
		\end{align}
	\end{subequations}
	
\begin{figure}[h]
		\centering
		\includegraphics[width=8.8cm]{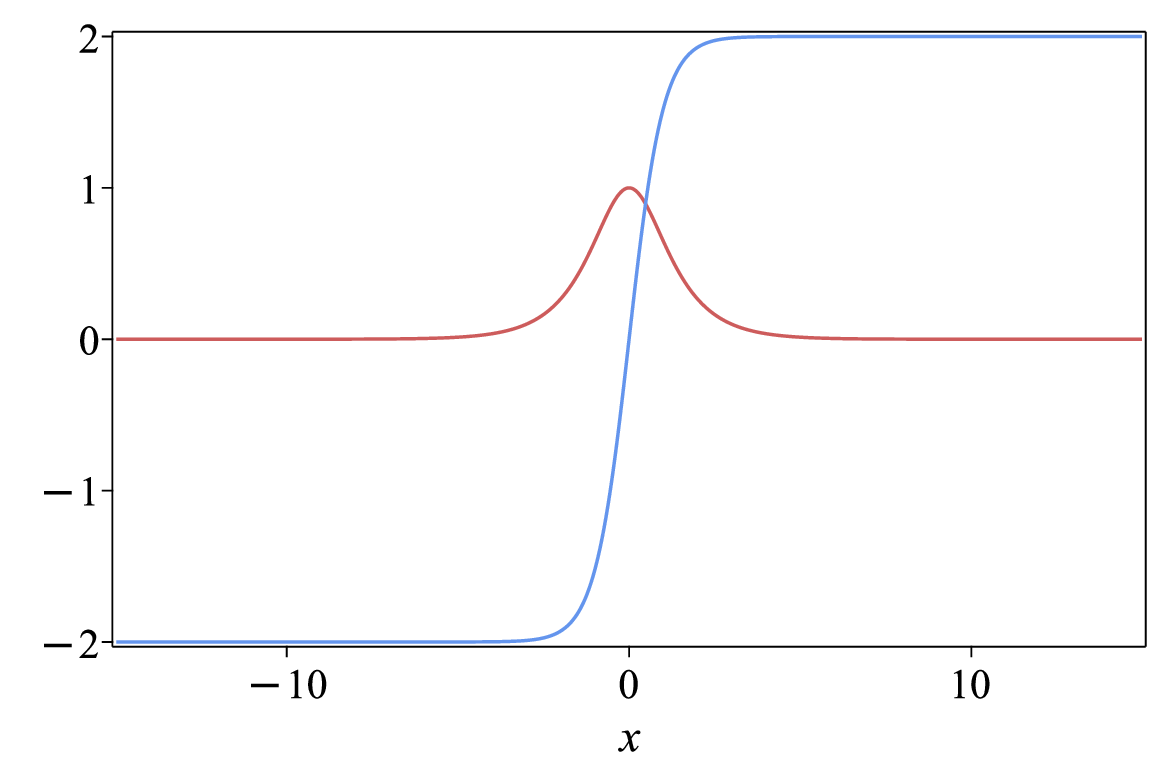}
		\caption{Solution $\phi=
			{\rm{sech}}(x)$ (red), $\chi=2\tanh(x)$ (blue) of~\eqref{FO1} with impurities~\eqref{impurityex3}, subject to boundary conditions~\eqref{bc2}.}
		\label{ex5fields}
	\end{figure}

	This topological solution, which is depicted in Fig.~\ref{ex5fields}, corresponds to a $\phi$ function that behaves as a lump peaked at the origin, together with a kink-like $\chi$ field with a zero at the same point. Taking the parity of these functions into account, it is not difficult to verify that both $W_{\phi}\pb{x}\phi$ and $W_{\chi}\pb{x}\chi$ are odd functions, which hence give a vanishing contribution to the energy. The solution is thus consistent with the result $E=\Delta W=0$ expected from saturation of the Bogomol'nyi bound. Interestingly, $W_{\phi}$ is in fact identically zero for this example, which means that the $\phi$ field only contributes to the energy density through the terms involving its derivative, which is completely determined by $\sigma_1$. This is also the reason $\phi$ is found to be an even function, even though the Lagrangian lacks the typical $\text{Z}_2$ symmetry. When combined, the form of $\sigma_1$ and the property $W_{\phi}=0$ ensure that all terms involving $\phi$ in $\LL$ are even functions, partially recovering the symmetry \emph{on shell}. On the other hand $W_{\chi}$ is nonzero, and in fact its on-shell value exactly cancels out the odd part of $\sigma_2$, thus explaining the fact that $\chi$ is an odd function.

	Since this BPS sector has no impurity-free analogue, it is perhaps important to ponder over the existence and eventual behavior of zero modes for the solution ~\eqref{sol3x3}. Intuition would perhaps point to the belief that normalizable solutions of the zero-mode equations should exist. After all, despite its zero energy, this solution does not appear fundamentally different from those of other topologically nontrivial sectors. \cmmnt{The solution $(\phi,\chi)=\left(\tanh\left(\frac{x}{2}), \sqrt{2}\sech\left(\frac{x}{2}\right)\right)\right)$, for example, can be derived in a different sector when $\sigma_i=0$. This solution possesses qualitative features quite similar to those seen in Fig.~\ref{ex5fields},  and is known to allow zero modes readily derivable from its orbit equation.} Despite these observations, our numerical investigations have proven unable to find another BPS solution consistent with conditions~\eqref{bc2}. A similar difficulty was encountered in attempts to solve the zero-mode equation obtained from~\eqref{sol3x3}, for which only exponentially growing, and thus unacceptable, modes have been found. These calculations seem to indicate the absence of zero modes in this sector, but this conjecture is difficult to demonstrate with certainty, as it is usually not easy to show a boundary value problem does not admit a solution. A similar situation was in fact investigated in Ref.~\cite{BLMPLB}, where the absence of zero modes for a topologically nontrivial impurity-domain-wall solution in two spatial dimensions has been formally demonstrated. 
	
	The possibility of a topologically nontrivial solution that does not allow zero-modes is a novel feature of the model, since such modes exist for all solutions of the model conceived in~\cite{BNRT}. This may have interesting consequences for some physical applications, specially those in which the scalar fields are used to generate internal structure in another field configuration, as is the case, for example, in some of the aforementioned brane applications of this model~\cite{BlochBrane, thick}. Normally, the internal structure generated by topological solutions of the model would be slowly modified by the rolling instabilities associated with zero-modes. The example just discussed illustrates the fact that these transformations may cease to be an issue if the model is doped with properly chosen impurities.

	These results can in fact be made more general, although that requires somewhat more complicated choices for the impurity functions. Indeed, one may verify that the impurities
	\begin{subequations}
		\begin{align}
			\begin{split}
				&\sigma_1(x) =\beta^2 \text{sech}^2(x) - \beta\text{sech}(x)\tanh(x), \\
				& \ \ \ +\frac{\tanh^2( x)}{4r}-1
			\end{split}\\
			&\text{and} \nonumber\\
			&\sigma_2(x) = \frac{2\text{sech}^2(x)+ \beta \text{sech}(x)\tanh(x)}{2\sqrt{r}}.
		\end{align}
	\end{subequations}
	lead to BPS equations that can be solved by 
	\begin{subequations}
		\begin{align}
			\phi(x) &=\beta\text{sech}(x), \\
			\chi(x) &= \frac{\tanh(x)}{\sqrt{r}}.
		\end{align}
	\end{subequations}

	The examples in this subsection are meant to illustrate one important possibility engendered by the models considered in this work: the creation of BPS sectors, in the sense that the addition of impurities to the homogeneous model has modified the field equations in such a way as to make Bogomol'nyi bound saturation possible in topological sectors which formerly allowed no BPS states. This achievement is owed to the fact that the impurity functions fundamentally modify the vacuum structure of the model, thus allowing for nontrivial zero energy solutions.
	
	\subsection{Fourth model}
	
\begin{figure}[h]
		\centering
		\includegraphics[width=8.8cm]{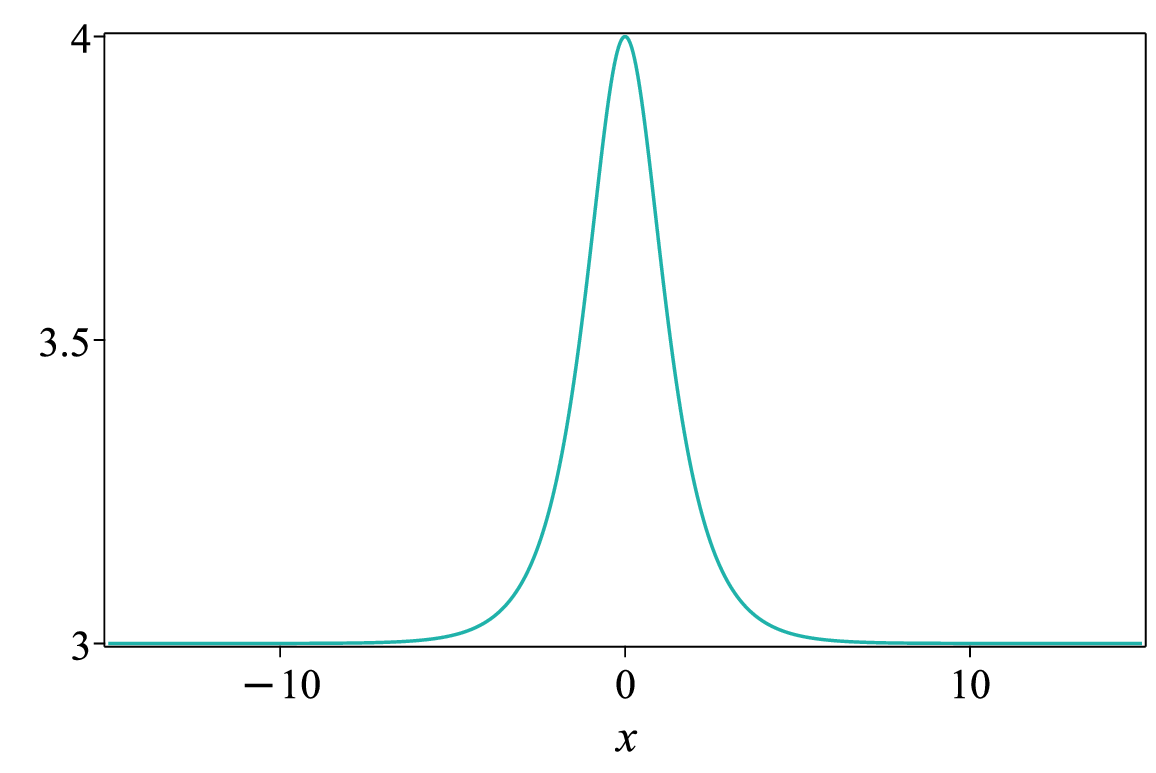}
		\caption{Impurity function $\sigma_1(x)$ given by equation~\eqref{impurityex4a}.}
		\label{sigmaex4}
	\end{figure}

	In the previous examples, we have dealt exclusively with localized impurities. As indicated in our earlier discussion in Section~\ref{gensec}, this need not be the case. Indeed, it is also possible to consider impurities which have arbitrary constants for their asymptotic limits, as long as~\eqref{boundary} are satisfied. In this section, we shall consider solutions engendered by one such impurity function. To this end, let
	\begin{subequations}\label{impurityex4}
		\begin{align}
			\sigma_1(x) &=3+\sech(x) \label{impurityex4a} \\
			\sigma_2(x) &=\frac{1}{1+x^2}.
		\end{align}
	\end{subequations}
	In Fig.~\ref{sigmaex4}, we show a plot of $\sigma_1$, while $\sigma_2$ was already depicted in Fig.~\ref{sigmaex1} from our first example.
	
\begin{figure}[H]
		\centering
		\includegraphics[width=8.8cm]{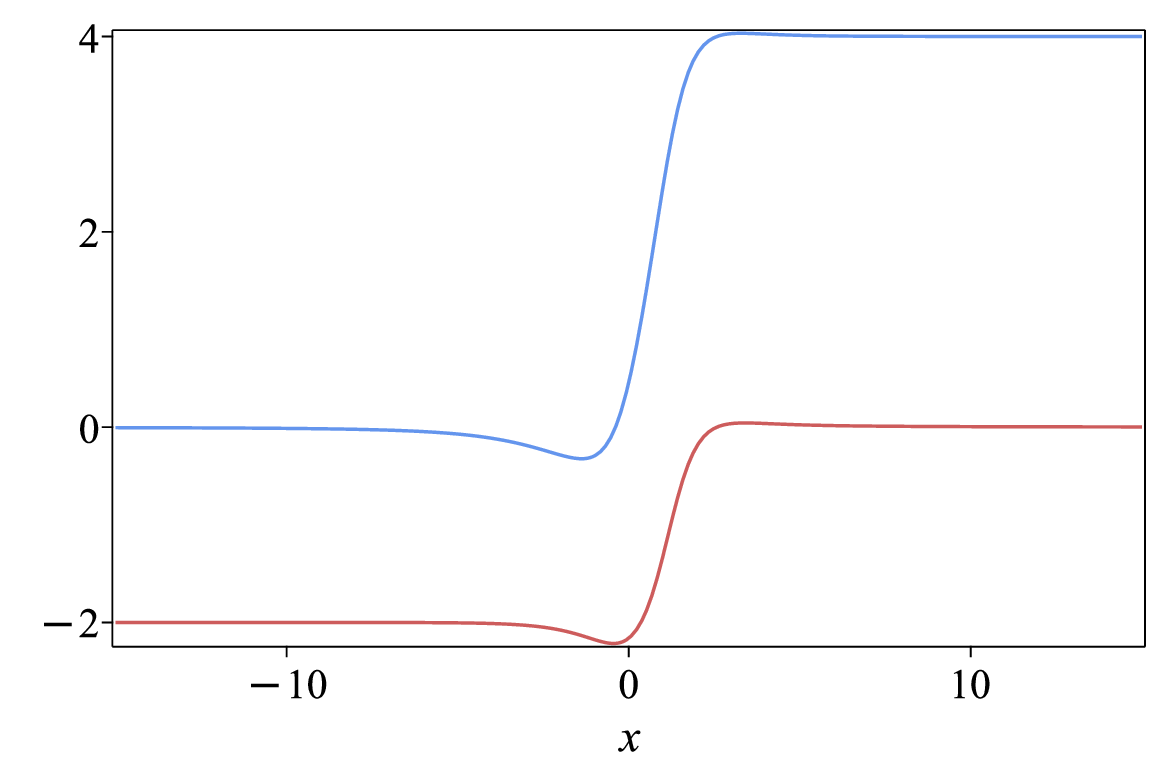}
		\caption{Solution $\phi$ (red), $\chi$ (blue) of~\eqref{FO1} with impurities~\eqref{impurityex4} and $r=1/4$.}
		\label{ex6fields}
	\end{figure}

	This pair of impurities may be distinguished from the functions previously considered by the fact that $\sigma_1$ tends to three, rather than zero, at plus and minus infinity. This asymptotic behavior generates a new set of boundary conditions for the problem, since the fields cannot, in this case, have the same limiting values as their impurity-free counterparts. One possibility is 
	\begin{subequations}\label{bc3}
		\begin{align}
			&\lim_{x\to_{-\infty}}\phi=-2, && \lim_{x\to_{-\infty}}\chi=0,\\
			&\lim_{x\to_{\infty}}\phi=0, && \  \lim_{x\to_{\infty}}\chi=2\sqrt{r},
		\end{align}
	\end{subequations}
	which corresponds to kink-like profiles for both $\phi$ and $\chi$. In particular, these fields tend, at large positive $x$, to trivial configurations which do not belong to the vacuum manifold of the $\LL_{\0}$ theory. The profile of these functions is depicted in Fig.~\ref{ex6fields}.

\begin{figure}[h]
		\centering
		\includegraphics[width=8.6cm, trim={1cm 0 0 0}]{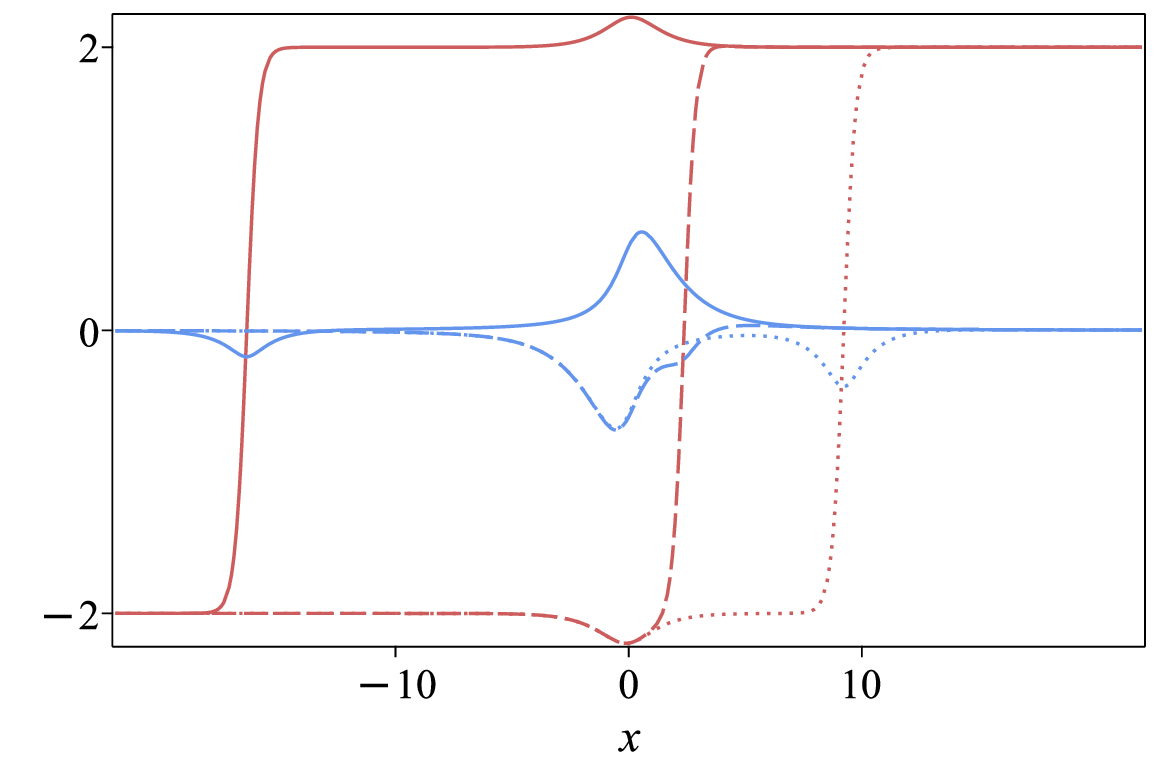}
		\caption{Solutions $\phi$ (red), $\chi$ (blue) of~\eqref{FO1} with impurities~\eqref{impurityex4} and $r=1/4$. $(\phi(0), \chi(0))$ are equal, with three significant digits, to $(2.21,0.593)$, $(-2.21,-0.612)$ and $(-2.21,0.592)$ for the solid, dashed and dotted-line solutions, respectively.}
		\label{ex7Afields}
	\end{figure}

\begin{figure}[h]
		\centering
		\includegraphics[width=8.5cm,trim={1cm 0 0 0}]{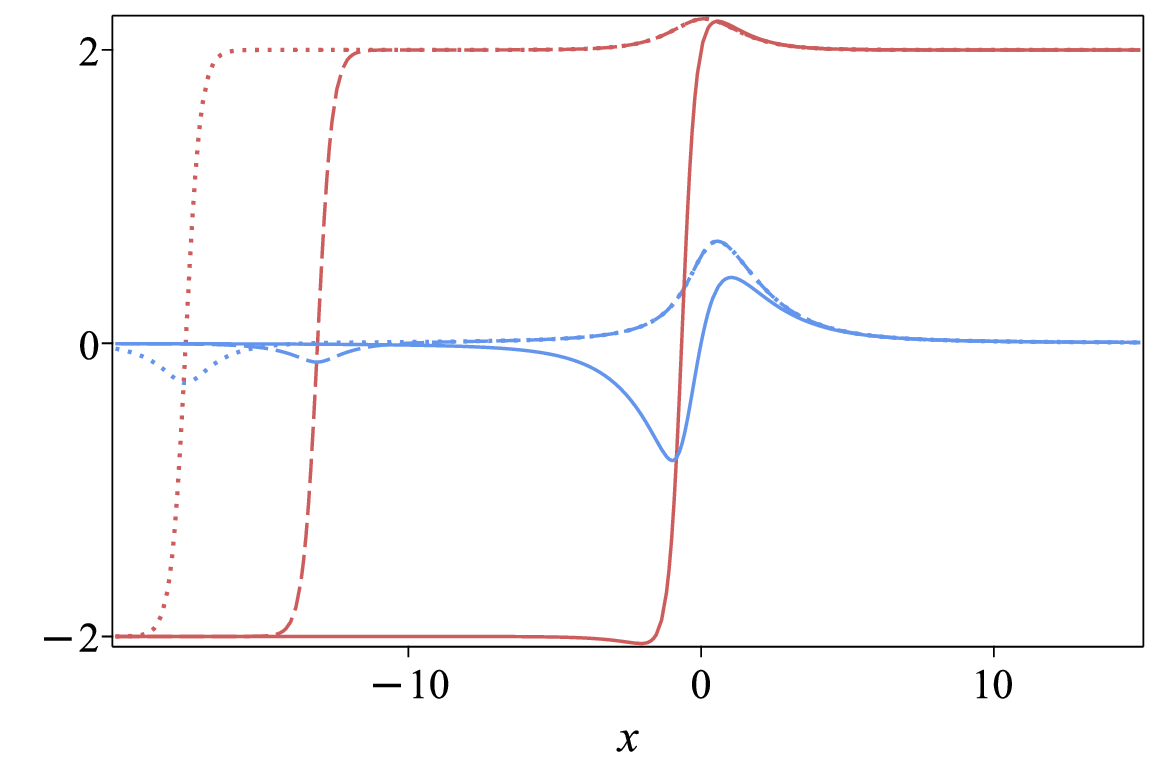}
		\caption{Solution $\phi$ (red), $\chi$ (blue) of~\eqref{FO1} with impurities given by~\eqref{impurityex4} and $r=1/4$. The values of $(\phi(0), \chi(0))$ rounded off to three digits are $(2.21,0.592)$,$(2.21,0.592)$ and $(2.00,0.000)$ for the dotted, dashed and solid-line solutions, respectively.}
		\label{ex7Bfields}
	\end{figure}

	Equations~\eqref{bc3} do not constitute the only set of acceptable, topologically nontrivial, boundary conditions enabled by the impurity functions~\eqref{impurityex4}. One may, for example, consider the problem defined by the conditions 
	\begin{subequations}\label{bc4}
		\begin{align}
			&\lim_{x\to_{-\infty}}\phi=-2, && \lim_{x\to_{-\infty}}\chi=0,\\
			&\lim_{x\to_{\infty}}\phi=2, && \  \lim_{x\to_{\infty}}\chi=0.
		\end{align}
	\end{subequations}
	Solutions have been found in the topological sector defined by these conditions. Here we have undercovered a very rich space made out of the many inequivalent BPS states within this topological sector specified by boundary conditions~\eqref{bc4}. In Figs.~\ref{ex7Afields} and~\ref{ex7Bfields} we depict a total of six solutions belonging to this topological sector. As the parameters used to numerically solve this boundary value problem are varied, the point of intersection between the functions slowly moves, and both the number and nature of critical points changes drastically. Since these solutions must be obtainable from each other via zero-mode perturbations, they can all be changed into each other after being affected by vanishingly small perturbations. Even the \qt{lump} seen, for example, in the solid blue line of Fig.~\ref{ex7Afields}, may be continuously deformed into the dashed-line configuration from the same Figure, thus suffering a change of concavity. It is also notable the extent to which the height (depth) of the maxima (minima) of the defects is controlled by the distance between their centers and the location of the impurities.

	
	\subsection{Topologically trivial solutions}\label{lump}
	
\begin{figure}[h]
		\centering
		\includegraphics[width=8.8cm,trim={1.2cm 0 0 0}]{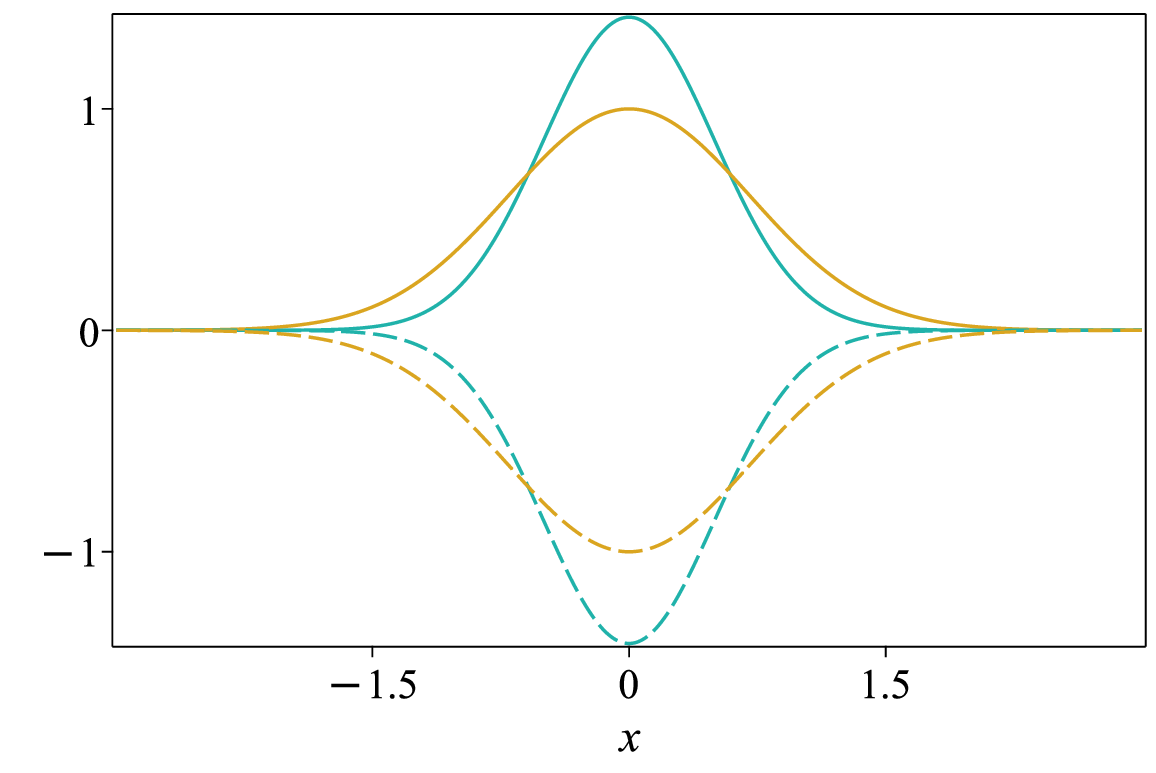}
		\caption{Gaussian impurity functions $\sqrt{2}e^{-2x^2}$ (solid green), $-\sqrt{2}e^{-2x^2}$ (dashed green), $e^{-x^2}$ (solid yellow) and $-e^{-x^2}$ (dashed yellow).}
		\label{sigmaex5}
	\end{figure}

	Thus far we have mostly concerned ourselves with topologically nontrivial configurations, a group to which belong all solutions from the previous subsections. In the standard theory the applicability of nontrivial configurations with zero topological charge is limited by the well-known instability issues, which fundamentally stem from the fact that any such solution belongs to a sector that is also populated by a trivial vacuum solution, to which it must decay. However, when one deals with inhomogeneous settings, stable solutions of this kind with nontrivial structure are perfectly possible, which follows immediately from the fact that the lowest energy in the vacuum sector ceases to be a trivial solution in the presence of impurities. \cmmnt{Indeed, it is clear that a homogeneous configuration $(\phi_{\0}, \chi_{\0})$ could never solve~\eqref{EL} for nonzero $\sigma_i$, so that even topologically trivial solutions of these equations must display some lump-like behavior or some kind of internal structure.} Among all static solutions of the Euler-Lagrange equations subject to nontopological boundary conditions, the stable configurations are selected by the requirement that they must solve equations~\eqref{FO} in the appropriate sector, so we proceed in analogous fashion to what we have developed before. With this in mind, let us now consider the boundary conditions
	\begin{subequations}\label{bc5}
		\begin{align}
			&\lim_{x\to_{-\infty}}\phi=1, && \lim_{x\to_{-\infty}}\chi=0,\\
			&\lim_{x\to_{\infty}}\phi=1, && \  \lim_{x\to_{\infty}}\chi=0,
		\end{align}
	\end{subequations}
	which pertain to a vacuum sector whose Bogomol'nyi energy is, of course, zero. In the impurity-free model, these boundary conditions correspond to a trivial BPS sector, in the sense that they only allow for the homogeneous solution $(\phi_{\0},\chi_{\0})=(1,0)$, whose fields never leave their vacuum values. We shall solve the BPS equations for these boundary conditions in a system doped by impurity functions of the Gaussian form, namely
	\begin{subequations}\label{impurityex5}
		\begin{align}
			\sigma_1(x)=&\alpha e^{-ax^2}, \label{impurityex5a}\\
			\sigma_2(x)=&\beta e^{-bx^2}, \label{impurityex5b}
		\end{align} 
	\end{subequations}
	where $\alpha$, $\beta$, $a$ and $b$ are again real constants, the later two of which are required to be positive to avoid infinite energy solutions in the theory. Gaussian functions are well-understood, and have often been used in investigations involving impurities. The profile of these functions  is illustrated in Fig.~\ref{sigmaex5}.

\begin{figure}[H]
		\centering
		\includegraphics[width=8.8cm]{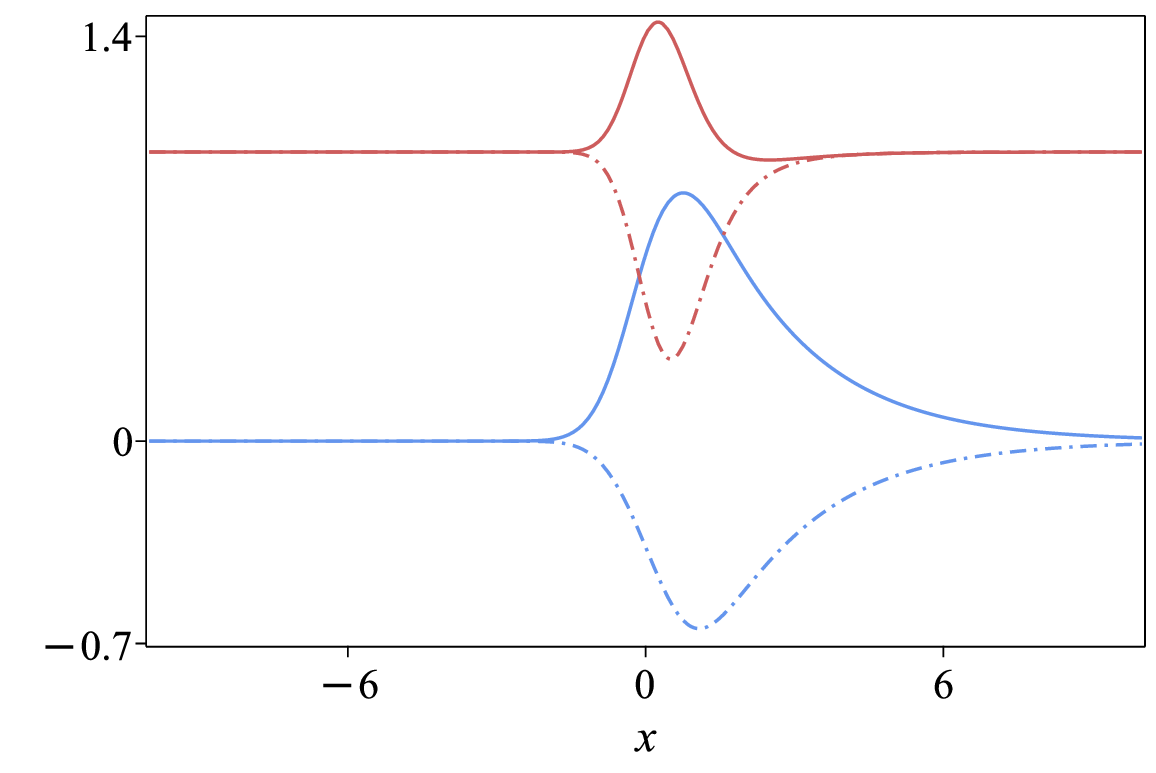}
		\caption{Solution $\phi$ (red), $\chi$ (blue) of~\eqref{FO1} with impurity functions given by $(\sigma_1,\sigma_2)=(\pm\sqrt{2}e^{-2x^2}, \pm e^{-x^2})$ and $r=1/4$. Solid lines specify the pair obtained from the upper (positive) signs, while the dashed-dotted solution corresponds to the lower ones.}
		\label{ex8fields}
	\end{figure}

	Equations~\eqref{FO1} have been solved numerically with the aforementioned impurities and boundary conditions. The results are displayed in Fig.~\ref{ex8fields}, where lump-like profiles for both fields are seen in the case of positive $\alpha, \beta$, while a similar solution with inverted concavity was obtained for negative values of these constants. In the standard theory, some topological sectors possess BPS solutions in which one of the fields is shown to have a lump profile, but it is not possible for both fields to be lump-like in a stable solution.
	
\begin{figure}[h]
		\centering
		\includegraphics[width=8.8cm]{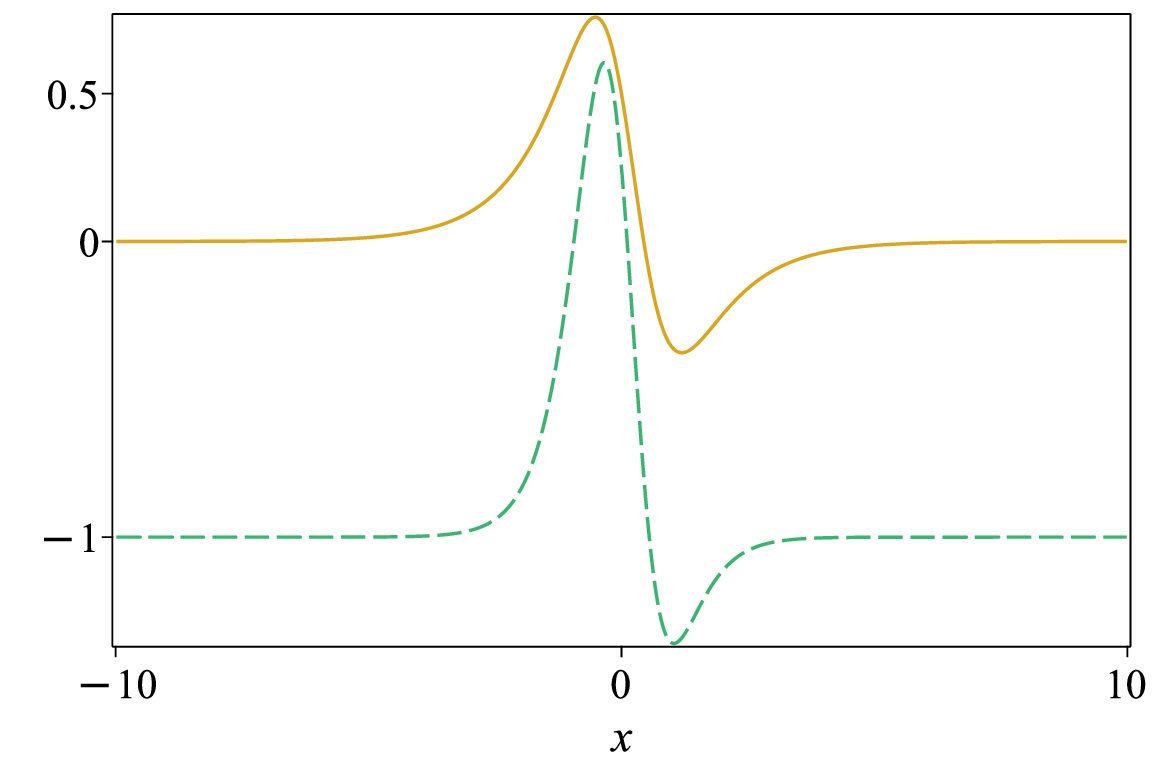}
		\caption{Impurity functions $\sigma_1$ (green dashed line), and $\sigma_2$ (yellow, solid line) from equations~\eqref{eq58}.}
		\label{sigmaex7}
	\end{figure}

\begin{figure}[h]
		\centering
		\includegraphics[width=8.8cm]{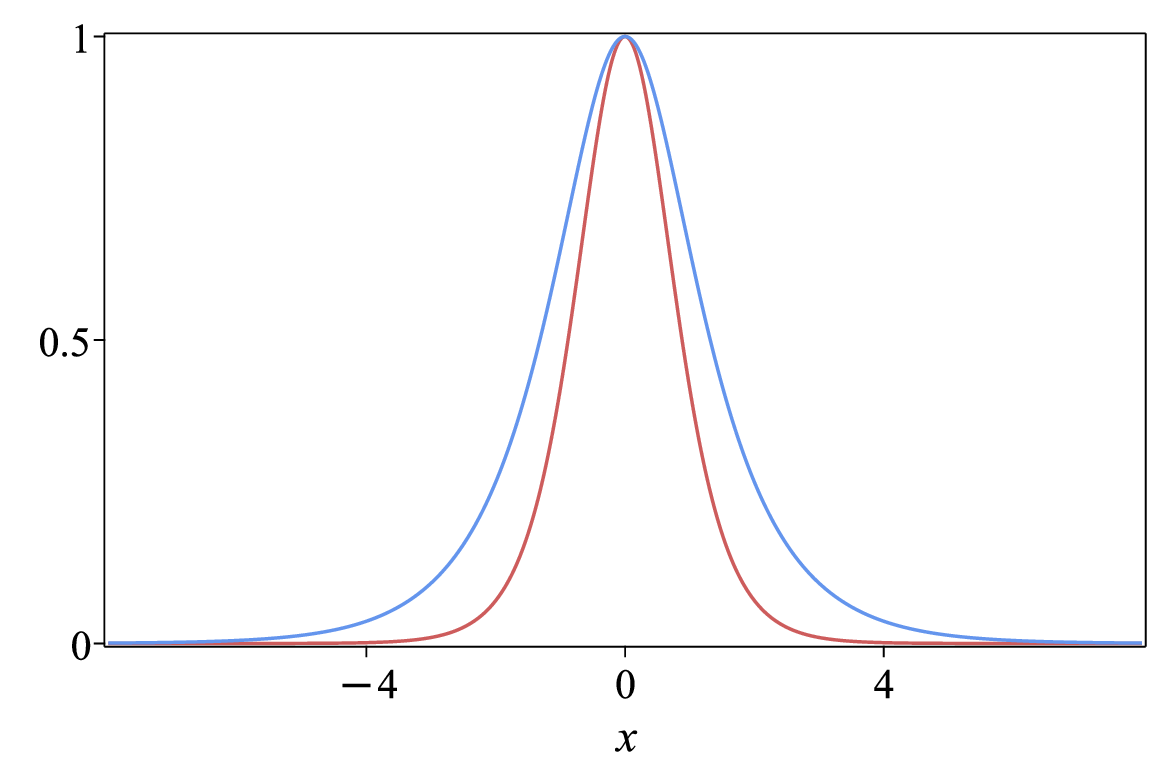}
		\caption{Solution $\phi=\rm{sech}^2(x)$ (red), $\chi=\rm{sech}(x)$ (blue) of~\eqref{FO1} with impurities~\eqref{eq58}, subject to boundary conditions~\eqref{bc5}.}
		\label{ex7fields}
	\end{figure}

	Since boundary conditions~\eqref{bc5} define a topologically trivial sector, one would not expect zero modes to exist for this solution. This is in fact what happens in the one-field theory, in which the symmetry transformations that give rise to kink zero-modes can be shown to leave the topologically trivial lump unchanged~\cite{AdamII}. Numerical analysis seems to indicate that the same is true for the solutions shown in Fig.~\ref{ex8fields}, as these boundary conditions appear to always lead to the same solution within numerical standards of accuracy. This result is in agreement with physical intuition, since one would not expect that the deformation effected by fixed impurity functions should give rise to infinitely many solutions. Instead of relying on numerical justification let us, however, propose a toy model, designed to produce lump-like solutions similar to the ones we just found, but chosen in such a way as to result in zero-mode equations that can be analytically solved in terms of special functions. To this end, consider the pair of impurities
	\begin{subequations}\label{eq58}
		\begin{align}
			\sigma_1 &= \operatorname{sech}(x)^2 \left(\frac{1}{4}+\operatorname{sech}(x)^2 - 2 \tanh(x)\right) -1, \\
			\sigma_2 &= \frac{\operatorname{sech}(x)^3}{2} - \operatorname{sech}(x) \tanh(x),
		\end{align}
	\end{subequations}
	whose profiles can be seen in Fig.~\ref{sigmaex7}. This choice gives rise to Bogomol'nyi equations that can be solved by the simple closed-form configurations
	\begin{align}\label{solutionex7}
		\phi(x)=\sech^2(x), && \chi(x)=\sech(x),
	\end{align}
	which display the sought-for lump-like profiles seen in Fig.~\ref{ex7fields}.

	\cmmnt{
		\begin{subequations}\label{zm}
			\begin{align}
				\frac{d}{dx} \etasc_1(x) &= -2 \operatorname{sech}(x)^2 \etasc_1(x) - \frac{\operatorname{sech}(x) \etasc_2(x)}{2}, \\
				\frac{d}{dx} \etasc_2(x) &= -\frac{\operatorname{sech}(x) \etasc_1(x)}{2} - \frac{\operatorname{sech}(x)^2 \etasc_2(x)}{2}.
			\end{align}
	\end{subequations}}

	We may now investigate the zero-mode equations related to this solution. To do this, one must substitute the fields~\eqref{solutionex7} into~\eqref{zeromodeBNRT}. This procedure results in first-order equations that one can integrate analytically to obtain a general solution that can be written in the form
	\begin{subequations}
		\begin{align}
			\etasc_{1}=C_1\psi_1 + C_2\psi_2, &&
			\etasc_{2}=C_1\zeta_1 + C_2\zeta_2 ,
		\end{align}
	\end{subequations}
	where
	\begin{widetext}
		\begin{align*}
			&\psi_1(x)=-\frac{\sech(x)}{e^{\frac{\tanh(x)}{2}}}\text{H}_c' \left( 3, -\frac{1}{2}, -\frac{1}{2}, -\frac{3}{2}, \frac{11}{8}, \frac{1}{2} + \frac{\text{tanh}(x)}{2} \right), && \zeta_1(x)=e^{-\frac{\tanh(x)}{2}}\text{H}_c \left( 3, -\frac{1}{2}, -\frac{1}{2}, -\frac{3}{2}, \frac{11}{8}, \frac{1}{2} + \frac{\text{tanh}(x)}{2} \right),
		\end{align*}
		\begin{align*}
			&\psi_2(x) =
			-\frac{\sech(x)}{e^{\frac{\tanh(x)}{2}}}\left[
			\sqrt{2\text{tanh}(x) + 2} \, \text{H}_c' \left( 3, \frac{1}{2}, -\frac{1}{2}, -\frac{3}{2}, \frac{11}{8}, \frac{1}{2} + \frac{\text{tanh}(x)}{2} \right) + 2 \, \frac{\text{H}_c \left( 3, \frac{1}{2}, -\frac{1}{2}, -\frac{3}{2}, \frac{11}{8}, \frac{1}{2} + \frac{\text{tanh}(x)}{2} \right)}{\sqrt{2\text{tanh}(x) + 2}}
			\right] ,\\
			&\text{and} \\
			&\zeta_2(x) = \text{H}_c \left( 3, \frac{1}{2}, -\frac{1}{2}, -\frac{3}{2}, \frac{11}{8}, \frac{1}{2} + \frac{\text{tanh}(x)}{2} \right)
			e^{-\frac{\text{tanh}(x)}{2}} \sqrt{2 + 2 \, \text{tanh}(x)} 
			\, .
		\end{align*}
	\end{widetext}	
	Here, $H_c(\alpha,\beta,\gamma,\eta,z)$ and  $H_c'(\alpha,\beta,\gamma,\eta,z)$ are respectively  the confluent form of the Heun function and its derivative~\cite{Heun}. 
	
\begin{figure}[h]
		\centering
		\includegraphics[width=8.8cm]{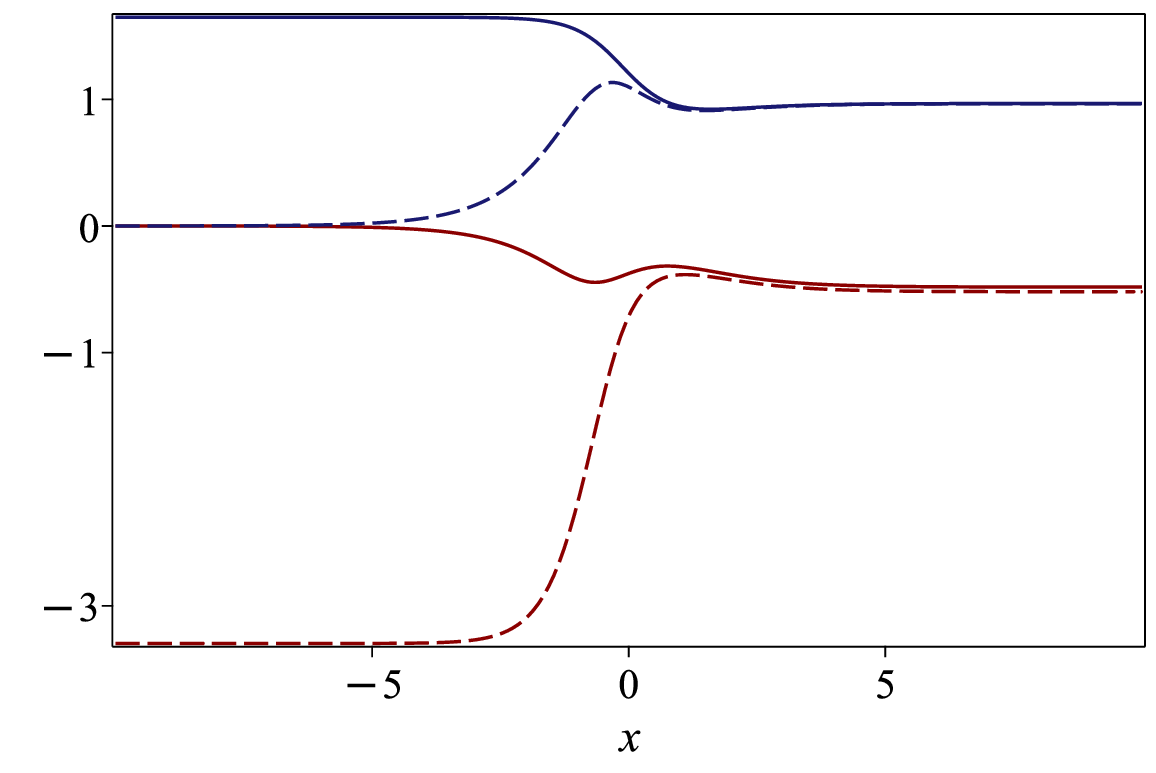}
		\caption{Linearly independent solutions of the zero-mode equation obtained from perturbations of~\eqref{solutionex7}. Here, $\psi_j$ and $\zeta_j$ are respectively represented in dark red and dark blue. Solid lines correspond to $j=1$, while dashed lines represent $j=2$.}
		\label{nozeromode}
	\end{figure}

	These functions can be implicitly defined in terms of the general solution of the differential equation~\cite{Decarreu}
	
	\begin{equation}
		\begin{split}
			\frac{d^2}{dz^2} Y(z) &- \frac{\left(-z^2 \alpha + (-\beta + \alpha - \gamma - 2)z + \beta + 1\right)}{z(z-1)} \frac{d}{dz} Y(z)\\ &-\left[ \frac{ (\beta + 1)\alpha + (-\gamma - 1)\beta - 2\eta - \gamma}{2z(z-1)} \right. \\
			&\ + \left.\frac{( \frac{\beta + \gamma}{2} +1)\alpha - \delta)}{z-1}\right]Y(z)=0, \nonumber
		\end{split}
	\end{equation}
	which has as its linearly independent solutions
	\begin{equation*}
		Y(x)=A H_c(\alpha,\beta,\gamma,\eta,z) + B z^{-\beta}H_c(\alpha,-\beta,\gamma,\eta,z).
	\end{equation*}
	For further information about confluent Heun functions, including its asymptotic behavior and expansions around their singularities, the reader is directed to~\cite{NISTII, Heunfunc, Heunasympt}, and references therein. These asymptotic results can be used to verify the behavior depicted in Fig.~\ref{nozeromode}, where it is seen that $(i)$ $\psi_1,\psi_2$ and $\zeta_1,\zeta_2$ tend respectively to the same nonzero values at infinity and $(ii)$ $\psi_2$, $\zeta_1$ do not tend to zero at minus infinity. Thus, it is easily seen that no nontrivial linear combination of these functions could ever go to zero at both limits of the domain. Indeed, $(i)$ can only be consistent with a zero limit if $C_1=-C_2$ which, by $(ii)$, means that $(\etasc_1,\etasc_2)$ go to zero at minus infinity if and only if $C_1=C_2=0$. This demonstrates that a normalizable zero mode cannot exist for boundary conditions~\eqref{bc5}. Thus, one must conclude that~\eqref{solutionex7} is the only solution that can saturate the Bogomol'nyi bound in this topological sector, and therefore the only stable solution in the static limit. These configurations can be viewed as deformations, performed by the impurity functions, in the trivial vacuum solution of the impurity-free theory, to which the fields~\eqref{solutionex7}  tend outside of a neighborhood of the $\sigma_i$, as consistency demands. Other solutions with different internal structure can, and in fact have been, found in this topological sector by directly solving the second-order equations. But, although qualitatively interesting, we shall not dwell on any such solutions because of their evident instability, which follows from the fact that their energy is higher than that of the BPS solution of this sector, to which they must thus unavoidably decay.
	
	\section{Conclusions and expected developments}\label{Conclusions}
	In this paper we have considered two-field configurations obtained from an impurity-generated deformation of a canonical Lagrangian $\LL_{\0}$. This deformation breaks the translational invariance of this theory, but the coupling between the impurity functions and the scalar fields is chosen in such aa way as to preserve part of its BPS sectors, in which stable configurations can be found.
	
	We have studied the general features of this model, finding both the Euler-Lagrange and the Bogomol'nyi equations for appropriate potentials. This discussion was followed by a stability analysis, where we have derived the stability Hamiltonian and the related eigenvalue problem, with particular emphasis in the zero-mode equations. We then considered specific examples, where it has been shown that the impurity-doped solutions possess many interesting features, such as a novel internal structure, which often includes properties not attainable in the impurity-free scenario (e.g., break of monotonicity in some sectors). Some discussions about the possibility of multiple solutions within a given topological sector, which are related to the existence of zero-modes, have also been made, and we have seen that the many acceptable BPS solutions from a sector may display highly nontrivial differences between themselves. Moreover, it has been shown that the addition of impurities may be used to create new topological sectors in the theory, and we have shown that it is also possible for doping to cause BPS solutions to emerge in a topological sector which did not support any solutions (trivial or not) in the impurity-free theory. This last situation is particularly interesting, and does not have any direct analogue in the one-field scenario, where the homogeneous theory always allows vacuum solutions in their zero-energy sectors.

	One promising venue of new developments is found when the homogeneous Lagrangian $\LL_{\0}$ is taken in the form
	\begin{equation}\label{L0gc}
		\begin{split}
			\LL_{\0}&=\frac12{P(\chi)\pu\phi\Pu\phi} +\frac12{Q(\phi)\pu\chi\Pu\chi}  \\
			& -\frac{W_\phi^2}{2P(\chi)}-\frac{W_\chi^2}{2Q(\phi)},
		\end{split}
	\end{equation}
	where $P(\chi)$ and $Q(\phi)$ are positive functions of the fields. This Lagrangian presents generalized  derivative kinetic and gradient terms of the form considered in~\cite{BLM2020}. In the particular case of separable superpotentials, the theory leads to solutions which are formally equivalent to geometrically constrained configurations generated by one of the fields, provided this field is held fixed. The extension of this work to the impurity-doped scenario that was investigated in the present work is in consideration for a future work in early stages of development, and appears to lead to interesting possibilities. Moreover,~\eqref{L0gc} is far from the only interesting noncanonical generalization in scalar field theories. Models requiring more general forms of the Lagrangian are also frequently studies, and have indeed found many important applications, with $K$-essence theories~\cite{KessenceI,KessenceII, KessenceIII}, $K-$defects~\cite{babichev, kadam, Andrews}, and cuscuton-like dynamics~\cite{Cuscuton, CuscutonII, CuscutonIII, CuscutonIV} being important examples that provide strong motivations for further investigations.
	
	Another possible development concerns the generalization of these results to the case of an arbitrary number of fields. This leads to Euler-Lagrange equations of the form~\eqref{fieldeq} and BPS configurations solving~\eqref{FOgen}. Given these equations and the fact that the entire analysis conducted in Section~\ref{stability} is valid for any number of fields, such a generalization should be straightforward, with expected results similar to the ones considered here. However, theories with three or more fields find some relevant applications in the impurity-free scenario (see, for example, Refs.\cite{ThreeFieldI,ThreeFieldII,ThreeFieldIII}), and those interested in such applications may find use in the generalized version of the theory. 
	
	Other possible generalizations include the extension to more than one spatial dimension. Theories with one real scalar field have been considered in an arbitrary number of dimensions, both in flat spacetimes~\cite{PLB22}, where it has been shown that the presence of impurities helps the circumvention of the usual scaling arguments, and in spherically symmetric spacetimes~\cite{BLM24}, yet none of these results has, to this date, been generalized to accommodate more than one field.
	
	Besides the extensions to new classes of models, there are also many aspects that are worth investigating in the currently considered theory. Although our examples and theoretical predictions include the features engendered by several types of inhomogeneities, other infinite possibilities do remain, and may lead to interesting results not yet accounted for. It may, for example, be possible to engender other types of asymptotic behavior, such as logarithmic or super-exponential tails~\cite{superexp}, or even total compactification of solutions, achieved if the field configurations meet their asymptotic values at finite distances of the origin, as is known to sometimes be possible in soliton theory~\cite{CompactonsI, CompactonsII}. The aforementioned properties would greatly impact the long-range interactions of the defects, and may thus be important for some physical applications. 
	
	There are also several theoretical issues we have not yet investigated, such as scattering processes, which have led to the discovery of some interesting phenomena in impurity-doped theories~\cite{SpectralWallsI,SpectralWallsII, PRD102}. Another issue that warrants further investigation is the spectral structure of the models we have considered. As indicated by the results we have presented, the existence and properties of solution of the stability equations are comparatively more complicated than in the one field theory. These equations may thus provide fertile ground for studies aiming to prove existence theorems, or for a qualitative analysis that may include investigations about higher excitation modes, which can play important roles in field dynamics. Finally,  attempts at a better understanding of the symmetry responsible for generating the zero modes of this theory, which would require an investigation along the lines of that developed in~\cite{AdamI}, may lead to interesting results, and to a deeper understanding of the model.
	
	\acknowledgements{This work is supported by the Brazilian agencies Conselho Nacional de Desenvolvimento Cient\'ifico e Tecnol\'ogico (CNPq), grants Nos. 402830/2023-7 (DB and MAM),  303469/2019-6 (DB), 306151/2022-7 (MAM) and  151204/2024-1 (MAL).}

\end{document}